\documentclass[10pt]{iopart}
\pdfoutput=1
\usepackage{graphicx}
\eqnobysec

\newcommand{\beq}{\begin{equation}}
\newcommand{\beqa}{\begin{eqnarray}}
\newcommand{\eeq}{\end{equation}}
\newcommand{\eeqa}{\end{eqnarray}}
\newcommand{\abs}[1]{\vert#1\vert}
\renewcommand{\e}{{\rm e}}
\newcommand{\eps}{\epsilon}
\newcommand{\epsr}{\eps_{\rm r}}
\newcommand{\g}{\gamma}
\newcommand{\half}{{\frac{1}{2}}}
\newcommand{\ii}{{\rm i}}
\renewcommand{\l}{\lambda}
\newcommand{\lra}{\leftrightarrow}
\renewcommand{\max}{{\rm max}}
\newcommand{\mean}[1]{\langle#1\rangle}
\newcommand{\s}{\sigma}
\newcommand{\stat}{{\rm stat}}
\newcommand{\w}{\widehat}
\newcommand{\E}{\Lambda}
\renewcommand{\H}{{\cal E}}
\newcommand{\Int}{\mathop{\rm Int}\nolimits}
\renewcommand{\Im}{\mathop{\rm Im}\nolimits}
\renewcommand{\Re}{\mathop{\rm Re}\nolimits}
\newcommand{\vecH}{{\bf H}}
\newcommand{\vecM}{{\bf M}}
\newcommand{\vecP}{{\bf P}}

\begin{document}

\title{Single-spin-flip dynamics of the Ising chain}

\author{C Godr\`eche and J M Luck}

\address{Institut de Physique Th\'eorique, Saclay, CEA and CNRS,
91191~Gif-sur-Yvette, France}

\begin{abstract}
We consider the most general single-spin-flip dynamics for the ferromagnetic
Ising chain with nearest-neighbour influence and spin reversal symmetry.
This dynamics is a two-parameter extension of Glauber dynamics corresponding
respectively to non-linearity and irreversibility.
The associated stationary measure is given by the usual Boltzmann-Gibbs
distribution for the ferromagnetic Hamiltonian of the chain.
We study the properties of this dynamics
both at infinite and at finite temperature, all over its parameter space, with
particular emphasis on special lines and points.
\end{abstract}

\eads{\mailto{claude.godreche@cea.fr},\mailto{jean-marc.luck@cea.fr}}

\vspace*{15pt}\address{\today}

\maketitle

\section{Introduction}
\label{sec:intro}

The one-dimensional kinetic Ising model introduced by Glauber in 1963~\cite{glauber}
is a prototypical model for the relaxation of a system towards thermal equilibrium.
The model evolves by non-conservative single-spin-flip dynamics,
with the requirement that the flipping rates fulfill the constraint of detailed
balance with respect to the ferromagnetic energy (or Hamiltonian)
\beq\label{eq:hamil}
\H=-\sum_n\s_n\s_{n+1},
\eeq
in dimensionless units, where $\s_n=\pm1$ are classical spins.
The detailed balance condition implies that the dynamics is reversible,
and hence that its stationary state is an equilibrium state.

It has recently been realized~\cite{gb09,cg11,cg13} that the most general
single-spin-flip dynamics
for the ferromagnetic Ising chain with nearest-neighbour influence and spin
reversal symmetry is a natural extension of Glauber dynamics
in two directions associated respectively to non-linearity and irreversibility.
The model introduced in~\cite{gb09,cg11,cg13} is defined by the rate
function~(\ref{eq:global1}) (or equivalently~(\ref{eq:global2})),
which are extensions of the Glauber rate functions~(\ref{glau})
and~(\ref{eq:global1}), where the parameters $\delta$ and $\eps$ take
arbitrary values in the triangular region depicted in
figure~\ref{fig:triangle}.
The stationary measure of this general model is that of the equilibrium
one-dimensional Ising model,
i.e., it is given by the usual Boltzmann-Gibbs distribution associated to the
ferromagnetic Hamiltonian~(\ref{eq:hamil}).

The aim of the present work is to pursue the study undertaken in~\cite{cg11},
where an exact analysis of the dynamical properties of the model along the line
$\delta=0$ of figures~\ref{fig:triangle} and \ref{fig:triangle2} was performed.
The present study is concerned with the properties of the model at a generic
point of the parameter space represented in figures~\ref{fig:triangle} and
\ref{fig:triangle2}, with particular emphasis on special lines and points.

The main focus will be on the dynamical behaviour of two-spin correlation functions
for a system started in either a random initial state or in a thermalized initial state.

\begin{itemize}

\item
In the first case all spin configurations are equally probable,
each spin $\s_n$ taking the values $\pm1$ with probability $1/2$ independently
of the others, hence the two-time correlation
\beq
C_n(0,t)=\mean{\s_0(0)\s_n(t)}
\eeq
describes the transient regime of relaxation of the system to stationarity.
We shall more particularly consider the autocorrelation
\beq
C(0,t)=C_0(0,t)=\mean{\s_0(0)\s_0(t)},
\eeq
which provides a measure of the overlap of the spin configuration of the system
at time $t$ with its random initial state.

\item
In the second case the system remains stationary during its evolution, hence
the correlation
\beq
C_{n,\stat}(t)=\mean{\s_0(0)\s_n(t)}
\eeq
gives a measure of the fluctuations of the system at stationarity.
We shall more particularly consider the autocorrelation
\beq
C_\stat(t)=C_{0,\stat}(t)=\mean{\s_0(0)\s_0(t)}.
\eeq

\end{itemize}
We shall also provide some insight in the behaviour of the equal-time correlation
\beq
C_n(t)=\mean{\s_0(t)\s_n(t)},
\label{eqts}
\eeq
for the system initially prepared in a random initial state and relaxing
towards stationarity.
At long times this correlation converges to
\beq
C_n=\mean{\s_0\s_n}=(\tanh\beta)^{\abs{n}},
\label{cstat}
\eeq
denoting by $\beta=1/T$ the inverse temperature, which is the well-known
expression of the correlation of the one-dimensional equilibrium Ising model~\cite{baxter}.

The setup of the paper is as follows.
Sections~\ref{sec:reminder1} and~\ref{sec:reminder} provide some prerequisite
knowledge on the model investigated in the present work.
In section~\ref{sec:reminder1} we give a reminder of the dynamical rules
defining the model~\cite{gb09,cg13}.
In section~\ref{sec:reminder} we give a reminder of some exact results~\cite{cg11}
on the dynamics of this model on the solvability line ($\delta=0$).
We emphasize the existence of an oscillatory relaxation regime beyond a
temperature-dependent threshold.
Section~\ref{sec:infinite} is the bulk of the present work.
It is devoted to novel aspects of the infinite-temperature
dynamics of the model, be it reversible or irreversible, beyond the solvability line.
To this end we use a wide range of methods
coming from various branches of statistical physics,
each of these approaches shedding some light onto the problem from another angle.
More specifically, we first use two general approaches,
time series expansions (section~\ref{timeseries})
and mapping of the dynamics onto a quantum spin chain (section~\ref{sec:quantum}),
before we analyze some special lines and points,
including in particular the solvability line ($\delta=0$) (section~\ref{linear}),
the reversibility line ($\eps=0$) (section~\ref{reversibleline}),
the SEP point ($\delta=-1$, $\eps=0$) (section~\ref{sec:sep}),
the dual SEP point ($\delta=1$, $\eps=0$) (section~\ref{sec:dualsep}) and
the ASEP (microcanonical) line ($\delta=-1$) (section~\ref{asepm}).
We also present some observations on the dynamical behaviour at a generic point
(section~\ref{generic}), based on numerical simulations.
Finally, an investigation of the spectra of the Markov matrix
(section~\ref{spectrum}) provides a useful alternative tool
to understand the qualitative features of the dynamics.
In section~\ref{finite} we investigate the main novel features
of the finite-temperature dynamics
which were absent in the infinite-temperature situation.
This includes a study of the relaxation rate along the reversibility line
(section~\ref{freversible})
and the existence of an oscillating regime of relaxation beyond a threshold
in the generic irreversible case (section~\ref{fgeneric}).
We then study the dynamics of the energy density of the model (section~\ref{etsc}),
and close with an investigation of two special points on the reversibility line
(KDH and Metropolis) (section~\ref{sec:kdh}).
We conclude by a brief discussion of our results in section~\ref{discussion}.

\section{A reminder on the rate function of the model}
\label{sec:reminder1}

In this section we give an overview of the dynamical rules of the model
introduced in~\cite{gb09,cg11,cg13}.\footnote{We refer the reader to those
references for more details,
in order to keep the reminder contained in this section succinct.}

\subsection{The rate function of the model}

We first recall that the most general rate function $w_n$
for single-spin-flip dynamics $(\s_n\to-\s_n)$ in continuous time
with spatially homogeneous nearest-neighbour influence
obeying both detailed balance and spin reversal symmetry reads~\cite{glauber}
\beq\label{glau}
\fl
w_n=
\frac{\alpha}{2}\Bigl(1
-\frac{\g}{2}(1+\delta)\s_n(\s_{n-1}+\s_{n+1})+\delta\,\s_{n-1}\s_{n+1}\Bigr)
\hspace{30pt}\quad(\textrm{Glauber}).
\eeq
This rate function depends on two free parameters:
$\alpha$ (which fixes the time scale) and~$\delta$.
The parameter $\g$ is related to inverse temperature $\beta$ by
\beq\label{eq:gam}
\g=\tanh 2\beta,
\eeq
as a consequence of detailed balance~\cite{glauber}.
The so-called Glauber model usually refers
to the case where the parameter $\delta$ is set to zero.
The success of Glauber model relies on its solvability,
in the sense that its time-dependent behaviour can be determined exactly.
All spin correlation functions of interest indeed obey linear evolution equations,
that can be solved by analytical means.
Whenever $\delta$ is non zero, the solvability of the model is lost,
in the sense that the hierarchy of evolution equations for spin correlations
does not close.

It has recently been shown~\cite{gb09,cg13} that the generalization
of~(\ref{glau}) to a rate function only satisfying the condition of
{\it global balance} reads
\beq\label{eq:global1}
\fl
w_n=
\frac{\alpha}{2}\Bigl(1+\eps\,\s_{n-1}\s_n-(\g(1+\delta)+\eps)\s_n\s_{n+1}
+\delta\,\s_{n-1}\s_{n+1}\Bigr)\quad(\textrm{Generic}).
\eeq
This rate function now depends on the additional parameter $\eps$.
From now on, we set the time unit by fixing $\alpha=1$.
Global balance is a weaker condition than detailed balance.
It only requests that the (unique) stationary state of the system
is given by the same Boltzmann-Gibbs measure as at equilibrium,
corresponding to the ferromagnetic Hamiltonian~(\ref{eq:hamil})
at fixed temperature $T$.
A dynamics only obeying global balance is generically irreversible,
and therefore generically leads to a nonequilibrium stationary state.
Global balance contains detailed balance (leading to an equilibrium state)
as a particular case.

In the present context, the condition of detailed balance corresponds to fixing
the parameter $\eps$ to the value
\beq\label{eq:rev}
\epsr=-\frac{\g}{2}(1+\delta)
\eeq
(where r stands for reversible), which yields~(\ref{glau}) back, now rewritten
as
\beq\label{glaub}
\fl
w_n=
\frac{1}{2}\Bigl(1
+\epsr\,\s_n(\s_{n-1}+\s_{n+1})+\delta\,\s_{n-1}\s_{n+1}\Bigr)\hspace{60pt}\quad(\textrm{Glauber}).
\eeq
Whenever the condition $\eps=\epsr$ is not satisfied, the difference
\beq\label{eta}
\eta=\eps-\epsr
\eeq
quantifies both the irreversibility of the dynamics and its left-right
asymmetry,
as can be seen on the new form of~(\ref{eq:global1}) now rewritten as
\beq\label{eq:global2}
\fl
w_n=
\frac{1}{2}\Bigl(1+(\epsr+\eta)\s_{n-1}\s_n+(\epsr-\eta)\s_n\s_{n+1}
+\delta\,\s_{n-1}\s_{n+1}\Bigr)\quad(\textrm{Generic})
\eeq
which is invariant under the simultaneous change of $\eta$ into $-\eta$
and exchange of left and right.
This property only holds for the Ising chain and does not extend to
higher-dimensional situations~\cite{cg13}.
It turns out that the expression~(\ref{eq:global1}) (or
equivalently~(\ref{eq:global2}))
not only satisfies the requirement of global balance once temperature is known,
but represents {\it the most general rate function for single-spin-flip dynamics
with spin reversal symmetry and nearest-neighbour influence}.
Equation~(\ref{eq:gam}) now becomes a condition fixing the
temperature~\cite{cg13}.

\subsection{Spin-flip moves}

The spin-flip moves associated with each environment are listed in
table~\ref{tab:fmoves}.
The expressions of the rates corresponding to these moves
in terms of the parameters $\delta$ and $\eta$ (arbitrary)
and~$\epsr$ (given by~(\ref{eq:rev})) are deduced from~(\ref{eq:global2}).
In the rightest column, the subscripts in the rates $w_{\s\s'}$ refer to the
two neighbours $\s$ and $\s'$ of a flipping $(+)$ spin.
The rates $w_{++}$ and $w_{--}$ (lines 1 and 4), which depend on $\delta$ and $\epsr$,
respectively correspond to the creation and the annihilation of a pair of domain walls,
with energy cost $\Delta\H=\pm4$.
The rates $w_{+-}$ and $w_{-+}$ (lines 2 and 3), which depend on $\delta$ and $\eta$,
correspond to the motion of the domain walls,
respectively to the left and to the right, with no cost in energy ($\Delta\H=0$).
In other words, the parameter $\eta=(w_{+-}-w_{-+})/2$
is only involved in the asymmetric motion of the domain walls.

An alternative description of the Ising chain
can be given in terms of the domain walls~\cite{racz}.
Defining occupation numbers living on the bonds of the chain, according to
\beq
\tau_n=\half(1-\s_n\s_{n+1})=0\hbox{ or }1,
\eeq
then, if $\tau_n=1$ (i.e., $\s_n\s_{n+1}=-1$),
there is a particle (domain wall) on the corresponding bond;
if $\tau_n=0$ (i.e., $\s_n\s_{n+1}=+1$),
there is a hole (i.e., no domain wall).
On a finite system of $N$ sites (and $N$ bonds), with periodic boundary conditions,
the total ferromagnetic energy $\H$ is related to the
number $M$ of particles by $\H=-N+2M$.
The reactions among particles corresponding to the four spin-flip moves appear
in the fourth column of table~\ref{tab:fmoves}.

\begin{table}[!ht]
\begin{center}
\begin{tabular}{|c|c|c|c|l|}
\hline
$\#$&spin-flip move&spin-flip move&reaction&rate\\
\hline
1&$+++\;\to\;+-+$&$---\;\to\;-+-$&$00\;\to\;11$&$w_{++}=\half(1+\delta)+\epsr$\\
2&$++-\;\to\;+--$&$--+\;\to\;-++$&$01\;\to\;10$&$w_{+-}=\half(1-\delta)+\eta$\\
3&$-++\;\to\;--+$&$+--\;\to\;++-$&$10\;\to\;01$&$w_{-+}=\half(1-\delta)-\eta$\\
4&$-+-\;\to\;---$&$+-+\;\to\;+++$&$11\;\to\;00$&$w_{--}=\half(1+\delta)-\epsr$\\
\hline
\end{tabular}
\caption{List of spin-flip moves,
with corresponding reactions among particles,
and expressions of the corresponding rates
in terms of the parameters $\delta$ and $\eta$.
The parameter $\epsr$ is related to $\delta$ and temperature
by~(\ref{eq:rev}).}
\label{tab:fmoves}
\end{center}
\end{table}

\begin{figure}[!ht]
\begin{center}
\includegraphics[angle=0,width=1\linewidth]{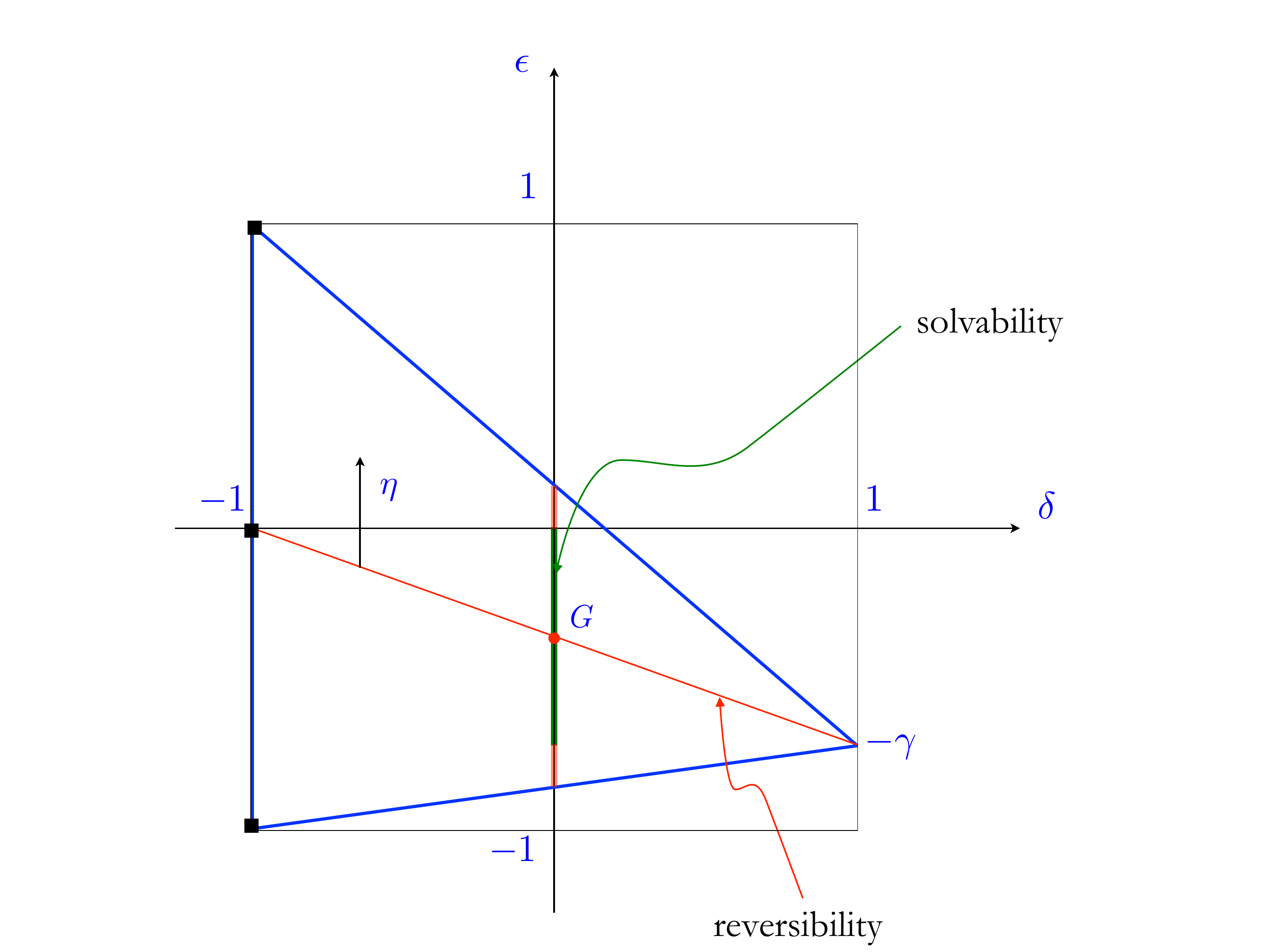}
\caption{Region of the $\delta$--$\eps$ plane where the dynamics obeys global balance
and all rates are positive,
with its two remarkable lines, the reversibility and solvability lines.
These lines intersect at point $G$, representing the Glauber model ($\delta=\eta=0$).
The irreversibility parameter $\eta$ is the
vertical coordinate of a generic point relatively to the reversibility line $\eps=\epsr$.
On the solvability line ($\delta=0$), correlation functions
exhibit an oscillatory temporal behaviour in the outer red segments,
i.e., for $\eps>0$ or $\eps<-\g$, or,
equivalently, for $\eta_0<\abs{\eta}<1/2$ (see~(\ref{eq:thresh})).
Black squares show the SEP point ($\delta=-1,\eps=0$) and the two TASEP
points ($\delta=-1,\eps=\pm1$) (After~\cite{cg13}.)}
\label{fig:triangle}
\end{center}
\end{figure}

\begin{figure}[!ht]
\begin{center}
\includegraphics[angle=0,width=1\linewidth]{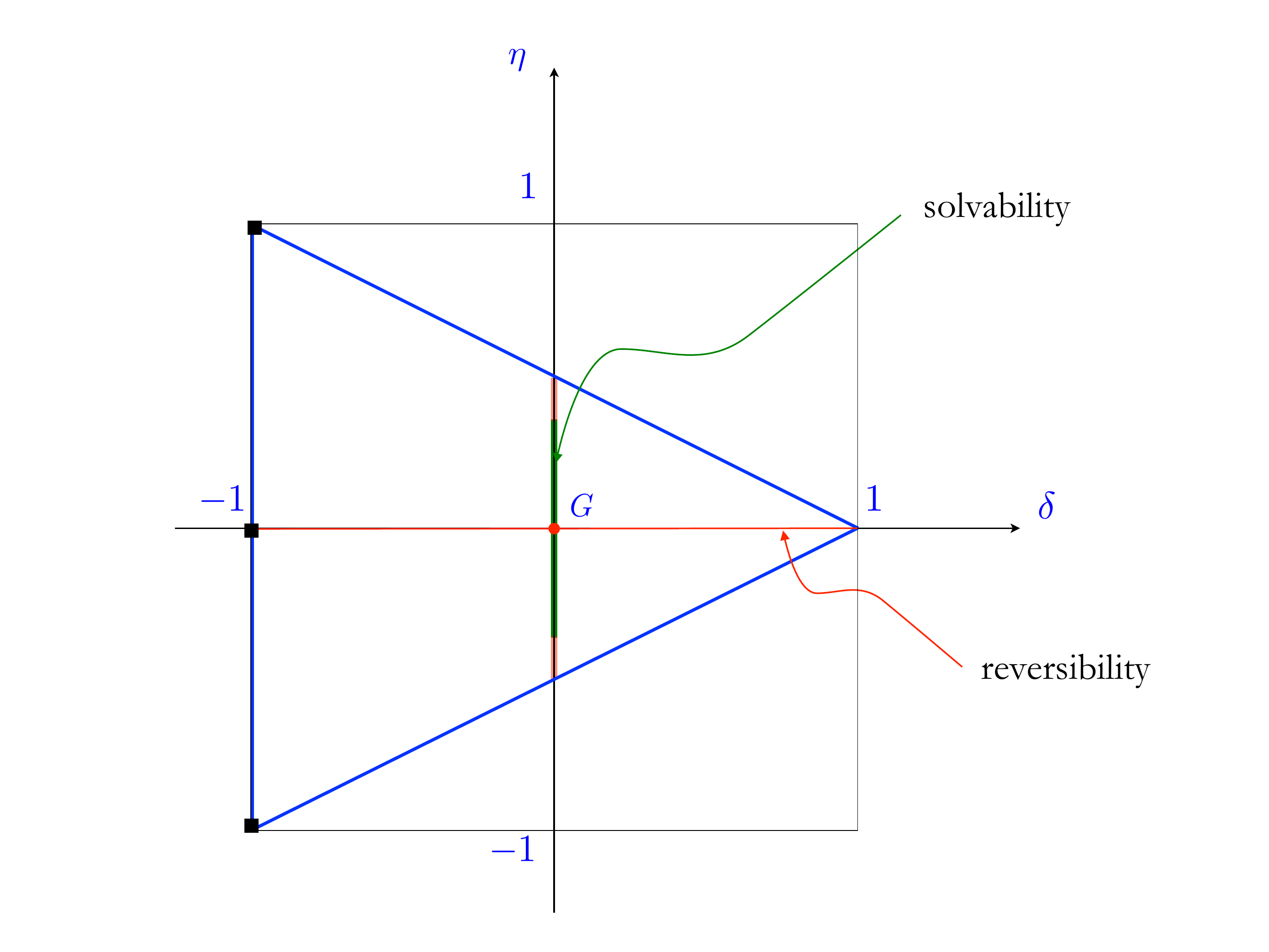}
\caption{Same as figure~\ref{fig:triangle}, represented in the $\delta$--$\eta$
plane.}
\label{fig:triangle2}
\end{center}
\end{figure}

The parameters $\delta$ and $\eps$ appearing in~(\ref{eq:global1}) are constrained
by the condition that all rates must be positive.
The parameter $\delta$ has to obey $\abs{\delta}\le1$,
while $\eta=\eps-\epsr$ has to obey $\abs{\eta}\le\eta_\max$, with
\beq
\eta_\max=\half(1-\delta).
\eeq
This defines a triangular region of the $\delta$--$\eps$ plane, depicted
in figure~\ref{fig:triangle}~\cite{cg13}.
Two lines on this plot are remarkable.
The first one is the {\it reversibility line} ($\eps=\epsr$, i.e., $\eta=0$),
which corresponds to all possible reversible dynamics.
The second one is the {\it solvability line} ($\delta=0$),
along which the dynamics is solvable,
in the sense that spin correlation functions obey a closed system of evolution equations.
These lines intersect at point~$G$, representing the Glauber model ($\delta=\eta=0$).
A detailed analysis of the dynamics along the solvability line,
beyond the Glauber model, was given in~\cite{cg11}
and will be recalled in section~\ref{sec:reminder}.
In figure~\ref{fig:triangle} the parameter $\eta$ represents the
vertical coordinate of a generic point relatively to the reversibility line $\eps=\epsr$.
A $\delta$--$\eta$ plot gives a symmetric and temperature-independent
representation of the triangle of all possible dynamics,
as depicted in~figure~\ref{fig:triangle2}.

For a generic point in the triangle
shown in figures~\ref{fig:triangle} and~\ref{fig:triangle2},
the rates obey the unique constraint
\beq\label{eq:db1}
\frac{w_{++}}{w_{--}}=\e^{-4\beta}=\frac{1-\g}{1+\g},
\eeq
which is one of the two conditions for detailed balance.
The other one,
\beq\label{eq:db2}
w_{+-}=w_{-+},
\eeq
only holds when the dynamics is symmetric or reversible ($\eta=0$).
The boundaries of the triangle correspond to the vanishing of one of the rates
given in table~\ref{tab:fmoves}.
The left vertical side of the triangle ($\delta=-1$) corresponds to
$w_{++}=w_{--}=0$, so that only the motions of domain walls are allowed.
Dynamics along this line will be hereafter referred to as microcanonical,
as they conserve the ferromagnetic energy.
On the two oblique lines one of the two conditions
$w_{+-}=0$ (lower side) and $w_{-+}=0$ (upper side) holds, i.e., $\eta=\mp\eta_\max$.
The point of intersection of these two oblique lines, to the right of the triangles,
corresponds to $w_{+-}=w_{-+}=0$, i.e., $\delta=1$ and $\eta=0$, or $\eps=-\epsr=-\g$.
The only allowed moves are the creation and the annihilation of pairs of domain walls,
with rates $w_{++}$ and $w_{--}$ respectively equal to $1\mp\g$.
More generally, the constant-energy moves, i.e., the motions of domain walls,
are more favoured (resp.~less favoured) than in the Glauber model for $\delta<0$
(resp.~$\delta>0$)~\cite{nemeth}.

\section{A reminder of the dynamics on the solvability line}
\label{sec:reminder}

The solvability line ($\delta=0$) is the natural extension to irreversible dynamics
of the finite-temperature Glauber model ($\delta=\eta=0$).
This line is parametrized by the irreversibility parameter $\eta=\eps-\epsr$,
with $\epsr=-\g/2$.
Although the dynamics is asymmetric and irreversible, it is still solvable~\cite{cg11},
in the sense that correlation functions obey a closed system of linear
evolution equations,
that can be solved analytically, just as for the Glauber
model~\cite{glauber,felderhof,gl2000}.
The outcomes are however surprisingly non trivial, as we recall below.

\subsection{Random initial state}

In the case where the system is started in a random initial state,
the two-time correlation $C_n(0,t)$ obeys the equation
\beq\label{eqdiff}
\frac{{\rm d}C_n(0,t)}{{\rm d}t}
=-C_n(0,t)-(\epsr-\eta)C_{n+1}(0,t)-(\epsr+\eta)C_{n-1}(0,t),
\eeq
with $C_n(0,0)=\delta_{n,0}$.
Introducing the Fourier transform
\beq
\w C(q,0,t)=\sum_n C_n(0,t)\e^{-\ii nq},
\eeq
we readily obtain
\beq
\w C(q,0,t)=\e^{-\Omega(q)t},
\eeq
with
\beq
\Omega(q)=1+2(\epsr\cos q-\ii\eta\sin q).
\label{fOmegaq}
\eeq

Let us first consider the case where $\abs{\eta}<\abs{\epsr}$.
We have
\beq\label{eq:Cn0t}
C_n(0,t)=\e^{-t}\left(\frac{\abs{\epsr}-\eta}{\abs{\epsr}+\eta}\right)^{n/2}
I_n\!\left(2t\sqrt{\epsr^2-\eta^2}\right),
\eeq
where the $I_n$ are the modified Bessel functions.
In particular the autocorrelation reads
\beq
C(0,t)=\e^{-t}I_0\!\left(2t\sqrt{\epsr^2-\eta^2}\right).
\label{fcweak}
\eeq
The above results were derived in~\cite{cg11},
where the ratio $\eta/\epsr$ is interpreted as a velocity and denoted by~$V$.
They demonstrate the existence of a threshold for the
irreversibility parameter~$\eta$, namely $\eta=\pm\eta_0$ (i.e., $V=\pm1$),
with
\beq\label{eq:thresh}
\eta_0=\abs{\epsr}=\frac{\g}{2}.
\eeq

As long as $\abs{\eta}<\eta_0$,
the correlation $C(0,t)$ decreases monotonically
and falls off exponentially, with the decay rate
\beq
\alpha_1=1-2\sqrt{\epsr^2-\eta^2}=1-\sqrt{\g^2-4\eta^2}.
\label{alpha1}
\eeq
This rate is minimal at the Glauber point ($\eta=0$), where it equals
\beq
\alpha_1=1-\g,
\label{fgap}
\eeq
and it increases towards 1 at threshold (see figure~\ref{fig:alpha}).

Right at the threshold values $\eta=\pm\eta_0$
(respectively corresponding to $\eps=0$ and $\eps=-\g$,
the two endpoints of the green segment on figures~\ref{fig:triangle}
and~\ref{fig:triangle2}),
the dynamics become totally asymmetric,
in the sense that the influence on the flipping spin comes from only
one of the neighbours~\cite{kun,gb09,cg11,cg13}, with respectively
\beq
w_n=\frac{1}{2}(1-\g\s_{n-1}\s_n)\quad\textrm{and}\quad
w_n=\frac{1}{2}(1-\g\s_n\s_{n+1}).
\eeq
Note that such a totally asymmetric dynamics can only occur if $\delta=0$,
since otherwise both neighbours always have an influence on the flipping spin.
For $\eta=-\eta_0$, the correlation $C_n(0,t)$ vanishes for $n<0$,
while for $n\ge0$ we have~\cite{cg11},
\beq\label{eq:v1}
C_n(0,t)=\frac{\e^{-t}(\gamma t)^n}{n!}.
\eeq
Likewise, for $\eta=\eta_0$, the same correlation vanishes for $n>0$,
while for $n\le0$ we have
\beq\label{eq:v2}
C_{n}(0,t)=\frac{\e^{-t}(\gamma t)^{-n}}{(-n)!}.
\eeq
The autocorrelation at the threshold, $C(0,t)=\e^{-t}$, is independent of temperature.

Beyond threshold ($\eta_0<\abs{\eta}<1/2$), i.e., in the outer red segments of
figures~\ref{fig:triangle} and~\ref{fig:triangle2},
the correlation $C(0,t)$ is given by the analytic continuation of~(\ref{fcweak}),
\beq
C(0,t)=\e^{-t}J_0\!\left(2t\sqrt{\eta^2-\epsr^2}\right),
\label{fcstrong}
\eeq
where $J_0$ is the usual Bessel function.
This correlation now exhibits a damped oscillatory behaviour,
in the form of asymptotically periodic oscillations,
multiplying an exponential decay with unit rate.
It vanishes for the first time at
\beq
t_1=\frac{j}{2\sqrt{\eta^2-\epsr^2}},
\label{ft1eta}
\eeq
where $j\approx2.404825$ is the first positive zero of $J_0$,
while subsequent zeros are asymptotically separated by the half-period
\beq
\Delta=\frac{\pi}{2\sqrt{\eta^2-\epsr^2}}.
\label{fdeleta}
\eeq
The time scales $t_1$ and $\Delta$ both diverge according to the same
inverse-square-root law as the threshold is approached ($\abs{\eta}\to\eta_0$).

The occurrence of an oscillatory behaviour beyond the threshold ($\abs{\eta}>\eta_0$)
can be given the following simple explanation.
Before threshold ($\abs{\eta}<\eta_0$),
both coefficients $-(\epsr+\eta)$ and $-(\epsr-\eta)$
which appear in~(\ref{eqdiff}) are positive.
The latter equation is therefore similar to a discrete diffusion equation,
and leads to monotonically decaying correlations.
Right at threshold ($\abs{\eta}=\eta_0$), one of the coefficients vanishes,
giving rise to a totally asymmetric dynamics and to the totally directed
correlations~(\ref{eq:v1})~and~(\ref{eq:v2}).
Finally, beyond threshold ($\abs{\eta}>\eta_0$),
the coefficients have opposite signs, giving rise to competing effects.
The interpretation of~(\ref{eqdiff}) as a discrete diffusion equation is lost.
The net outcome of these competing terms is the occurrence of damped oscillations.

\subsection{Thermalized initial state}
\label{sec:linthe}

The case where the system is started in a thermalized initial state
yields a richer behaviour, with two successive thresholds.

The correlation $C_{n,\stat}(0,t)$,
which accounts for the fluctuations of the system in the stationary state,
still obeys~(\ref{eqdiff}), albeit with the initial condition
$C_{n,\stat}(0,0)=v^{\abs{n}}$ (see~(\ref{cstat})), with $v=\tanh\beta$, so
that $\gamma=2v/(1+v^2)$, hence
\beq
\w C_\stat(q,0)=\frac{1-v^2}{1-2v\cos q+v^2}.
\label{structure}
\eeq
The linearity of the differential equation~(\ref{eqdiff}) ensures that
\beq
\w C_\stat(q,t)=\w C_\stat(q,0)\e^{-\Omega(q)t}.
\eeq
In other words, $C_{n,\stat}(0,t)$ is given by the spatial convolution
\beq
C_{n,\stat}(0,t)=\e^{-t}\sum_mv^{\abs{n-m}}
\left(\frac{\abs{\epsr}-\eta}{\abs{\epsr}+\eta}\right)^{m/2}
I_m\!\left(2t\sqrt{\epsr^2-\eta^2}\right).
\eeq
We have in particular
\beq
C_\stat(t)=\e^{-t}\sum_mv^{\abs{m}}
\left(\frac{\abs{\epsr}-\eta}{\abs{\epsr}+\eta}\right)^{m/2}
I_m\!\left(2t\sqrt{\epsr^2-\eta^2}\right).
\eeq
This correlation falls off exponentially whenever the irreversibility parameter
$\eta$ is less than the above threshold ($\abs{\eta}<\eta_0$).
The corresponding decay rate however takes two different values in the following regimes.

\begin{itemize}

\item {\it Regime~I},
corresponding to a weak violation of the equilibrium fluctuation-dissipation
theorem~\cite{cg11}, takes place for $\abs{\eta}<\eta_c$, with
\beq\label{eq:etac}
\eta_c=\frac{\g}{2}\sqrt{1-\g^2}.
\eeq
Here, $C_\stat(t)$ exhibits the same decay rate
as $C(0,t)$ (see~(\ref{alpha1})), i.e.,
\beq
\alpha_1=1-\sqrt{\g^2-4\eta^2}.
\eeq

\item {\it Regime~II},
corresponding to a strong violation of the fluctuation-dissipation theo\-rem~\cite{cg11},
takes place for $\eta_c<\abs{\eta}<\eta_0$.
Here, the decay rate of $C_\stat(t)$ reads
\beq
\alpha_2=\frac{2\abs{\eta}}{\g}\sqrt{1-\g^2}.
\eeq
We have $\alpha_2<\alpha_1$.

\end{itemize}

\begin{figure}[!ht]
\begin{center}
\includegraphics[angle=0,width=.7\linewidth]{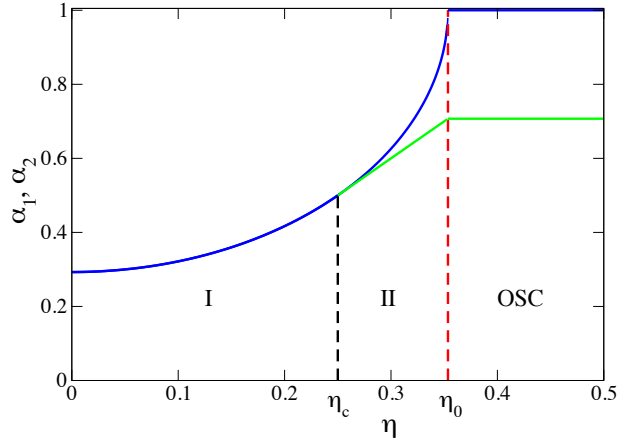}
\caption{Relaxation rates $\alpha_{1,2}$ against $\eta$
for $C(0,t)$ and $C_\stat(t)$.
In region~I,~$\alpha_1$ holds for both correlation functions (in blue).
In region~II, $\alpha_1$ holds for $C(0,t)$ (upper curve in blue), while
$\alpha_2$ holds for $C_\stat(t)$ (lower curve in green).
The black dotted vertical line is located at the critical value $\eta_c$, which
marks a bifurcation between two regimes of violation of the
equilibrium fluctuation-dissipation theorem.
The red dotted vertical line is located at the threshold value~$\eta_0$ beyond
which both correlation functions oscillate.
Temperature is chosen such that $\g=1/\sqrt{2}$.}
\label{fig:alpha}
\end{center}
\end{figure}

Both rates match at $\eta=\eta_c$, as well as their derivatives with respect to $\eta$,
i.e.,
\beq
\alpha_1=\alpha_2=1-\g^2,
\quad\frac{{\rm d}\alpha_1}{{\rm d}\eta}=\frac{{\rm d}\alpha_2}{{\rm
d}\eta}=\frac{2}{\g}\sqrt{1-\g^2}
\quad(\eta=\eta_c).
\eeq
Figure~\ref{fig:alpha} shows the $\eta$ dependence of the decay rates
$\alpha_{1}$ and $\alpha_{2}$.
Temperature is chosen in such a way that $\eta_c$ assumes its maximal value $\eta_c=1/4$.
This occurs for $\g=1/\sqrt{2}$, where we have
$\alpha_1(\eta_c)=\alpha_2(\eta_c)=1/2$, $\eta_0=1/(2\sqrt{2})$
and $\alpha_2(\eta_0)=1/\sqrt{2}$.

\begin{figure}[!ht]
\begin{center}
\includegraphics[angle=0,width=.7\linewidth]{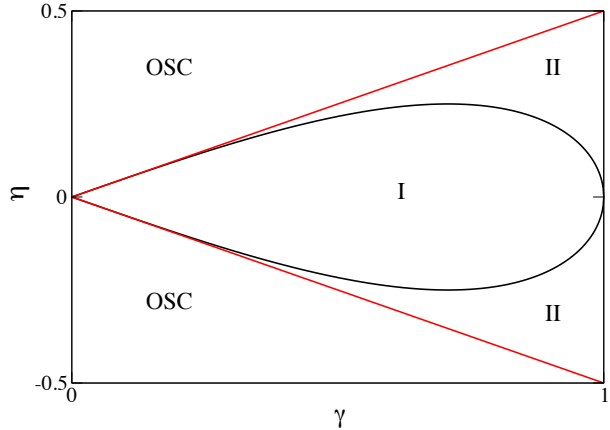}
\caption{\small
Dynamical phase diagram at finite temperature
in the $\g$--$\eta$ plane for the correlation $C(0,t)$ and
$C_\stat(t)$ on the solvability line ($\delta=0$).
Black curves: $\eta=\pm\eta_c$ (see~(\ref{eq:etac})).
Red lines: $\eta=\pm\eta_0$ (see~(\ref{eq:thresh})).
The threshold values occurring in figure~\ref{fig:alpha} are recovered
for $\g=1/\sqrt{2}$.}
\label{regimes}
\end{center}
\end{figure}

The behaviour of the correlation functions
$C(0,t)$ (random initial state) and $C_\stat(t)$ (thermalized initial state)
can be summarised as a phase diagram in the $\g$--$\eta$ plane,
shown in figure~\ref{regimes}.
In Regime~I ($\abs{\eta}<\eta_c$),
both correlations fall off monotonically to zero and share the same decay rate $\alpha_1$.
In Regime~II ($\eta_c<\abs{\eta}<\eta_0$),
the correlation functions still fall off monotonically to zero,
albeit with different decay rates,
namely~$\alpha_1$ for $C(0,t)$ and $\alpha_2<\alpha_1$ for $C_\stat(t)$.
Finally, in the regions marked OSC ($\abs{\eta}>\eta_0$),
both correlation functions exhibit oscillations.

In contrast, the equal-time correlation $C_n(t)$ does not depend {\it at all}
on the irreversibility parameter $\eta$~\cite{cg11}.
This is a symmetry of the solvability line,
which is (weakly) violated for generic values of the parameter $\delta$
(see section~\ref{etsc}).

We refer the reader to~\cite{cg11} for more details on the results mentioned
above and for further investigations on two-time observables.
A parallel study has been performed both for the spherical model~\cite{glv} and
for the two-dimensional Ising model~\cite{pleim2014}.

\section{Infinite temperature}
\label{sec:infinite}

We now address new aspects of the dynamics of the model, beyond the solvability line.
In this section, which is the bulk of the present work,
we focus our attention onto the case of infinite-temperature dynamics.
In spite of its simplicity, this situation already encompasses
most of the novel dynamical features of the generic model.
The rate~(\ref{eq:global1}) becomes
\beq
w_n=\half\Bigl(1+\eps\,\s_n(\s_{n-1}-\s_{n+1})+\delta\,\s_{n-1}\s_{n+1}\Bigr),
\label{infrate}
\eeq
where the two free parameters $\delta$ and $\eps$ respectively encode
the non-linearity and the irreversibility of the dynamics.
The parameter $\eps$ is indeed identical to $\eta$ at infinite temperature,
thus figure~\ref{fig:triangle} degenerates into figure~\ref{fig:triangle2}.
It has to obey $\abs{\eps}\le\eps_\max$, with
\beq
\eps_\max=\eta_\max=\half(1-\delta).
\label{epsmax}
\eeq

The dynamics defined by~(\ref{infrate}) is generically irreversible
and gives birth to a nonequilibrium stationary state,
except on the line $\eps=0$, where it is reversible and yields an equilibrium state.
This infinite-temperature stationary state
is the random state where all spin configurations are equally probable.
In other words, a random initial state is already thermalized.
The correlation functions $C(0,t)$ and $C_\stat(t)$ are therefore identical.

For the infinite-temperature Glauber model ($\delta=0$, $\eps=0$),
the spins flip with constant rate $1/2$,
and thus remain independent of each other in the course of time.
The two parameters of the model, $\delta$ and $\eps$,
deform this dynamics in a non-trivial fashion.
The very simple statics of the model
indeed does not preclude the occurrence of interesting dynamical features,
which are investigated hereafter by means of a variety of techniques.
We first use two general approaches, time series expansions (section~\ref{timeseries})
and mapping of the dynamics onto a quantum spin chain (section~\ref{sec:quantum}),
before we analyze some special lines and points,
including in particular the solvability line ($\delta=0$) (section~\ref{linear}),
the reversibility line ($\eps=0$) (section~\ref{reversibleline}),
the SEP point ($\delta=-1$, $\eps=0$) (section~\ref{sec:sep}),
the dual SEP point ($\delta=1$, $\eps=0$) (section~\ref{sec:dualsep}) and
the ASEP (microcanonical) line ($\delta=-1$) (section~\ref{asepm}).
We end up with some observations on the dynamical behaviour at a generic point
(section~\ref{generic}), based on numerical simulations,
and with an investigation of the spectra of the Markov matrix
of the dynamics (section~\ref{spectrum}).

\subsection{Time series expansion}
\label{timeseries}

In order to apprehend the role of the parameters $\delta$ and $\eps$,
a first approach consists in expanding the correlation $C(0,t)$
as a power series in time $t$.
This technique is presented in full detail in~\cite{timeseries}.
In the present context, it will prove useful in identifying various symmetries
of the dynamics.

Consider the correlation $\mean{\s_0(0)P_A(t)}$, where
\beq
P_A(t)=\prod_{n\in A}\s_n(t)
\eeq
is the product of the spins of an arbitrary finite set $A$ of sites.
This correlation is non-vanishing only if the size $\abs{A}$
of the set $A$ is an odd integer.
It obeys the linear differential equation
\beq
\frac{\rm d}{{\rm d}t}\mean{\s_0(0)P_A(t)}=-2\sum_{n\in
A}\mean{\s_0(0)w_n(t)P_A(t)}.
\eeq
The simplest of these equations, corresponding to $A=\{0\}$, reads explicitly
\beqa
\frac{\rm d}{{\rm d}t}\mean{\s_0(0)\s_0(t)}&=&-2\mean{\s_0(0)w_0(t)\s_0(t)}
\nonumber\\
&=&-\mean{\s_0(0)\s_0(t)}-\delta\mean{\s_0(0)\s_{-1}(t)\s_0(t)\s_1(t)}
\nonumber\\
&&+\eps\mean{\s_0(0)\s_1(t)}-\eps\mean{\s_0(0)\s_{-1}(t)}.
\eeqa
By considering larger and larger sets,
such as $A=\{-1\}$, $\{1\}$, $\{-1,0,1\}$, and so on,
and taking averages over the random initial state,
one can systematically derive the coefficients
\beq
a_k=(-1)^k\frac{{\rm d}^k}{{\rm d}t^k}\mean{\s_0(0)\s_0(t)}_{t=0}
\eeq
of the time series expansion of the correlation of interest:
\beq
C(0,t)=\sum_{k\ge0}a_k\frac{(-t)^k}{k!}.
\label{cseries}
\eeq
A similar expansion can be derived for any other local observable.

Symbolic routines run on the computer
allow one to derive up to 20 coefficients of the above expansion.
The first few of them read
\beqa
a_0=1,
\nonumber\\
a_1=1,
\nonumber\\
a_2=1+\delta^2-2\eps^2,
\nonumber\\
a_3=1+5\delta^2-6\eps^2,
\nonumber\\
a_4=1+18\delta^2+3\delta^4-(12+8\delta^2)\eps^2+6\eps^4,
\nonumber\\
a_5=1+58\delta^2+31\delta^4-(20+64\delta^2+4\delta^3)\eps^2+30\eps^4.
\label{ak}
\eeqa
For the infinite-temperature Glauber model ($\delta=\eps=0$),
we have $C(0,t)=\e^{-t}$ (see~(\ref{idt})) so all the coefficients $a_k$ equal unity.
Generically the $a_k$ are polynomials in $\delta$ and $\eps$,
whose degree grows linearly with the order $k$.

\subsection{Mapping onto a quantum spin chain}
\label{sec:quantum}

Another technique which is very commonly used
in investigations of kinetic Ising models~\cite{siggia,henkel1,henkel2,schutz}
consists in mapping the dynamics of a chain of classical spins
onto the statics of a quantum chain of spin operators in the spin-$1/2$ representation,
\beq
S_n^x=\half\s_n^x,\quad
S_n^y=\half\s_n^y,\quad
S_n^z=\half\s_n^z,
\eeq
where $\s_n^x$, $\s_n^y$, $\s_n^z$, are Pauli matrices acting at site
$n$.\footnote{This notation for the Pauli matrices should not be confused
with the notation $\s_n$ for the classical Ising spin sitting at site $n$.}

There are in general several ways of mapping either the dynamics of a classical
spin chain or a reaction-diffusion system onto a quantum Hamiltonian $\vecH$.
The review by Sch\"utz~\cite{schutz} presents a systematic way of doing so.
Following this route, we obtain in the present case
\beq
\fl
\vecH=\half\sum_n\Bigl(1-\s_n^x\s_{n+1}^x
+\delta(\s_n^y\s_{n+1}^y+\s_n^z\s_{n+1}^z)
+\ii\eps(\s_n^y\s_{n+1}^x-\s_n^x\s_{n+1}^y)\Bigr).
\label{ham}
\eeq

On the infinite-temperature reversibility line ($\eps=0$),
$\vecH$ coincides with the Hamiltonian
of the XXZ (anisotropic Heisenberg) quantum spin chain~\cite{yy,baxter},
with~$\delta$ being the anisotropy parameter.
This spin chain is known to be integrable in the usual sense
(existence of an infinity of conservation laws, applicability of Bethe Ansatz techniques).
This property will be exploited in section~\ref{reversibleline}.

Whenever the dynamics is irreversible ($\eps\ne0$),~(\ref{ham}) is
the Hamiltonian of the asymmetric XXZ chain, investigated in~\cite{aa}.
The latter Hamiltonian is still integrable, albeit non-Hermitian.
It has complex spectrum in general (see section~\ref{spectrum}).
For $\delta=-1$, i.e., along the ASEP line, ferromagnetic interactions become isotropic.
The corresponding Hamiltonian~\cite{gs,hs}
provides the basis for investigations of the ASEP
by Bethe Ansatz techniques (see~\cite{gmr} for a review).

\subsection{Solvability line ($\delta=0$)}
\label{linear}

In this section we mention briefly how the results concerning the solvability line,
recalled in section~\ref{sec:reminder}, simplify at infinite temperature.
Equation~(\ref{eqdiff}) becomes
\beq
\frac{{\rm d}C_n(0,t)}{{\rm d}t}=-C_n(0,t)+\eps(C_{n+1}(0,t)-C_{n-1}(0,t)),
\eeq
with the initial condition $C_n(0,0)=\delta_{n,0}$.
We thus obtain, in Fourier space, $\w C(q,t)=\e^{-\Omega(q)t}$, with
\beq
\Omega(q)=1-2\ii\eps\sin q,
\label{Omegaq}
\eeq
and finally $C_n(0,t)=\e^{-t}J_n(2\eps t)$.
In particular the autocorrelation reads $C(0,t)=\e^{-t}J_0(2\eps t)$.
In the reversible case,
i.e., for the infinite-temperature Glauber model ($\delta=\eps=0$),
we are facing a dynamics of independent spins with rate 1/2, hence
\beq
C(0,t)=\e^{-t}.
\label{idt}
\eeq
As soon as the dynamics is irreversible ($\eps\ne0$),
the correlation $C(0,t)$ exhibits an exponential decay with unit rate,
modulated by asymptotically periodic oscillations.
The dynamics is indeed always in its oscillatory regime,
because the threshold $\eta_0$ vanishes at infinite temperature.
The first zero of $C(0,t)$ occurs at time
\beq
t_1=\frac{j}{2\abs{\eps}},
\label{t1eps}
\eeq
where $j\approx2.404825$ is the first positive zero of $J_0$,
while subsequent zeros are asymptotically separated by the half-period
\beq
\Delta=\frac{\pi}{2\abs{\eps}}.
\label{deleps}
\eeq

\subsection{Reversibility line ($\eps=0$)}
\label{reversibleline}

The dynamics is reversible on the line $\eps=0$,
where the second detailed balance condition~(\ref{eq:db2}) holds.
This infinite-temperature reversible model has the following peculiarity.
The coeffi\-cients $a_k$ listed in~(\ref{ak})
appear to involve only even powers of the non-linearity parameter~$\delta$.
This suggests that the correlation $C(0,t)$ is symmetric under the change
$\delta\to-\delta$.
This is indeed an exact dynamical symmetry, which can be demonstrated as follows.
Define new spin variables~$\w\s_n$,
obtained from $\s_n$ by flipping every second pair of spins,
according to\footnote{$\Int(x)$ denotes the integer part of $x$,
i.e., the largest integer less than or equal to $x$.}
\beq
\w\s_n=(-1)^{\Int(n/2)}\s_n.
\label{ws}
\eeq
This construction was already used by N\'emeth~\cite{nemeth}.
Consider now a given spin flip,
expressed both in the original variables $\s_n$ and in the new variables $\w\s_n$.
If the $\s_n$ flip is of type~1 or 4 (see table~\ref{tab:fmoves}),
the $\w\s_n$ flip is of type 2 or 3, and vice versa.
In the infinite-temperature reversible case,
flips of types 1 and 4 have the rate $(1+\delta)/2$,
while flips of types 2 and~3 have the rate $(1-\delta)/2$.
This explains the observed symmetry.
The latter is broken as soon as the dynamics is irreversible ($\eps\ne0$).
This is testified by the presence of the term $4\delta^3\eps^2$
in the expression~(\ref{ak}) of $a_5$.

The spectrum of relaxation times of the model can be, at least in principle,
extracted from the corresponding quantum Hamiltonian $\vecH$.
In particular the relaxation rate $\alpha_1$,
characterizing the exponential fall-off of the correlation
\beq
C(0,t)\sim\e^{-\alpha_1 t},
\eeq
coincides with the spectral gap of $\vecH$ in the thermodynamic limit
and in the appropriate sector.
For the reversible model,
where $\vecH$ is the integrable Hamiltonian of the XXZ spin chain,
the gap in the relevant magnetic sector can be read off from~\cite{jb}.
In our units, it reads
\beq
\alpha_1=\sqrt{1-\delta^2}.
\label{gap}
\eeq
We have thus found an exact expression
for the relaxation rate of the reversible model at infinite temperature.

For the infinite-temperature Glauber model ($\delta=0$), we have $\alpha_1=1$,
in agreement with the exponential decay $C(0,t)=\e^{-t}$
of the correlation at all times (see~(\ref{idt})).
More interestingly, the above relaxation rate vanishes with a square-root
singularity at the endpoints of the reversibility line ($\delta\to\pm1$).
The properties of the dynamics at these points are addressed in
sections~\ref{sec:sep} and~\ref{sec:dualsep}.

\subsection{SEP point ($\delta=-1$, $\eps=0$)}
\label{sec:sep}

At this point, the dynamics looks particularly simple, as we have
\beq
w_{++}=w_{--}=0,\quad w_{+-}=w_{-+}=1.
\eeq
This dynamics was already considered by several authors~\cite{achiam,spohn}.
In terms of the spin variables,
it conserves the total energy $\H$ of the ferromagnetic model.
It may therefore be referred to as a microcanonical dynamics.
In terms of the particles representing domain walls,
the dynamics conserves the total number $M$ of particles.
It coincides with
the dynamics of the symmetric exclusion process (SEP)~\cite{schutz,liggett},
where the reversible diffusive moves $01\lra10$ occur with unit rate.
The dynamics at this point is entirely independent of temperature,
in agreement with its microcanonical character.

The SEP point is one of the endpoints of the reversibility line,
where the relaxation rate $\alpha_1$ vanishes (see~(\ref{gap})),
pointing toward a sub-exponential decay of the correlation $C(0,t)$.
The time series for the latter quantity
involves coefficients which are pure numbers:
\beqa
a_0&=&1,\quad
a_1=1,\quad
a_2=2,\quad
a_3=6,\quad
a_4=22,\quad
a_5=90,\quad
\nonumber\\
a_6&=&396,\quad
a_7=1848,\quad
a_8=9108,\quad
a_9=47400,\ \dots
\eeqa
These numbers are not listed in the OEIS~\cite{OEIS}.
It would be interesting to give them a combinatorial interpretation.

The spin correlation $C(0,t)$ can be recast in the language of the SEP as follows.
The spin~$\s_0$ flips each time a particle crosses the origin.
Let $Q(t)$ be the net number of particles which cross the origin
during a lapse of time of duration~$t$
(counted positively if moving to the right, and negatively if moving to the left)
in the stationary state of the SEP characterised by a particle density $\rho$
(a random initial spin state corresponds to~$\rho=1/2$).
We have
\beq
C(0,t)=\mean{(-1)^{Q(t)}}=\mean{\e^{\ii\pi Q(t)}}.
\label{cqiden}
\eeq
After a long time $t$, $Q(t)$ will be typically large,
and hence approximately given by $Q(t)\approx\rho R(t)$,
where $R(t)$ is the random position at time $t$ of the particle
which was the first to the right of the origin at time $t=0$, say.
The distribution of $R(t)$ has been studied by several authors~\cite{qt}.
The bulk of this distribution is known to be asymptotically a centered Gaussian,
with a variance growing as
\beq
\mean{R^2(t)}\approx\frac{2(1-\rho)}{\rho}\sqrt\frac{t}{\pi}.
\eeq
Taking this result literally,
forgetting about the discrete nature of particles,
we obtain the rough estimate
\beq
C(0,t)\sim\exp\left(-\frac{\pi^2}{2}\rho^2\mean{R^2(t)}\right)
\sim\exp\left(-\rho(1-\rho)\sqrt{\pi^3t}\right).
\eeq
A stretched exponential decay of the correlation $C(0,t)$, of the form
\beq
C(0,t)\sim\exp\left(-A(\rho)\sqrt{t}\right),
\label{carho}
\eeq
has indeed been predicted in~\cite{spohn},
where the amplitude $A(\rho)$ is expressed in terms of the solution
of a variational problem.
A quantitative prediction for this amplitude,
\beq
A(\rho)=\frac{1}{\sqrt\pi}\sum_{k\ge1}\frac{(4\rho(1-\rho))^k}{k^{3/2}},
\label{adg}
\eeq
can be read off from the work by Derrida and Gerschenfeld~\cite{dg},
involving the Bethe Ansatz and the use of results by Tracy and Widom.

Coming back to the Ising chain,
a random initial state corresponds to a particle density $\rho=1/2$,
where the amplitude $A(\rho)$ takes its maximal value
\beq
A(1/2)=\frac{\zeta(3/2)}{\sqrt\pi}\approx1.473874,
\label{amax}
\eeq
where $\zeta$ is the Riemann zeta function.
Interestingly enough, the particle density $\rho=1/2$ appears
as a singular point, around which the amplitude $A(\rho)$ has a triangular shape:
\beq
A(\rho)\approx A(1/2)-4\bigl\vert\rho-1/2\bigr\vert.
\eeq

\subsection{Dual SEP point ($\delta=1$, $\eps=0$)}
\label{sec:dualsep}

This point, where $w_{++}=w_{--}=1$ and $w_{+-}=w_{-+}=0$,
is the other endpoint of the infinite-temperature reversibility line.
In terms of the spin variables $\s_n$,
all moves which change the ferromagnetic energy~$\H$,
i.e., moves of types 1 and 4 (see table~\ref{tab:fmoves}),
are equally allowed and take place with unit rate.
In terms of the particles~$\tau_n$ representing domain walls,
the dynamics consists in the pair creation and annihilation
reactions $00\lra11$ with unit rate.
This dynamics has a simple alternative description.
In terms of the new (dual) particles
\beq
\w\tau_n=\half(1-\w\s_n\w\s_{n+1})=\left\{\matrix{
\tau_n\hfill & (n\hbox{ even}),\cr
1-\tau_n & (n\hbox{ odd})\hfill}\right.
\label{wtau}
\eeq
(see~(\ref{ws})), the dynamics is again that of a SEP.

The symmetry related to the transformations~(\ref{ws}) and~(\ref{wtau})
of spins and particles exchanges the SEP and dual SEP points.
In particular the correlation $C(0,t)$ is identical at both points.
We recall that this symmetry only holds in the infinite-temperature case.

\subsection{ASEP (microcanonical) line ($\delta=-1$)}
\label{asepm}

Along this whole line, we have $w_{++}=w_{--}=0$,
therefore the dynamics is microcanonical.
In terms of the particles representing domain walls,
the dynamics consists in the moves $01\to10$ and $10\to01$
with respective rates $1+\eps$ and $1-\eps$.
Particles go preferentially to the left for $\eps>0$ and to the right for $\eps<0$.
This dynamics is that of the asymmetric exclusion process (ASEP).
It is invariant under the simultaneous change of $\eps$ into $-\eps$
and exchange of left and right.
The reversible dynamics of the SEP point is recovered when $\eps=0$.
The endpoints $\eps=\pm1$ correspond to the totally asymmetric exclusion
process (TASEP)~\cite{schutz,liggett}.

The time series of the correlation $C(0,t)$ at both TASEP points
again involves coefficients which are positive integers,
\beqa
a_0&=&1,\quad
a_1=1,\quad
a_2=0,\quad
a_3=0,\quad
a_4=8,\quad
a_5=40,\quad
\nonumber\\
a_6&=&136,\quad
a_7=392,\quad
a_8=1032,\quad
a_9=3912,\ \dots
\eeqa

For arbitrary values of $\eps$,
the correlation $C(0,t)$ is still related by the iden\-tity~(\ref{cqiden})
to the net number~$Q(t)$ of particles which cross the origin
during a lapse of time of duration~$t$ in the stationary state of the ASEP
with a particle density $\rho$.
A random initial spin state again corresponds to~$\rho=1/2$.
The distribution of $Q(t)$ is however no longer Gaussian,
as was the case for the SEP.
It is indeed governed by the non-linear Kardar-Parisi-Zhang (KPZ) theory~\cite{kpz}.
The mean value of~$Q(t)$ is given by the mean stationary current,
i.e., $\mean{Q(t)}=Jt$, with $J=-2\rho(1-\rho)\eps$ in our units.
The magnitude of the fluctuations of $Q(t)$ around its mean value
depends on the way it is measured.
For a given particle and a given configuration of the system at time $t=0$,
typical fluctuations scale as $(\eps^2t)^{1/3}$.
Their statistics are by now well characterised~\cite{kpzres}:
they involve one of the universal laws originally discovered by Tracy and Widom
in random matrix theory.
The situation is however made more intricate
by the fact that intrinsic fluctuations are masked by statistical ones,
as soon as they are averaged either over different particles
and/or over different configurations of the system at time $t=0$~\cite{kpzfls}.
To the best of our knowledge, no analogue of the results~(\ref{carho}),~(\ref{adg})
has been derived for the ASEP so far.

We have investigated the correlation $C(0,t)$ by means of numerical simulations.
It is observed to exhibit a stretched exponential law of the form
\beq
C(0,t)\sim\exp\left(-A\sqrt{t}\right),
\label{caeps}
\eeq
i.e., the same functional form as the exact result~(\ref{carho})
which holds at the symmetric SEP point ($\eps=0$).
The amplitude $A$ depends continuously on the irreversibility parameter $\eps$.
The decay law~(\ref{caeps}) is modulated by oscillations,
whose half-period~$\Delta$ also depends on $\eps$ (see figure~\ref{asep} below).
The exponent 1/3 characterising the anomalous scaling of intrinsic fluctuations
in one-dimensional KPZ theory does not enter~(\ref{caeps}).
This feature, which may seem surprising at first sight,
is certainly related to the above discussion on the nature of the fluctuations.
It would be very desirable to obtain a derivation of our conjectured
result~(\ref{caeps}) from the vast body of knowledge on KPZ theory.

\begin{figure}[!ht]
\begin{center}
\includegraphics[angle=0,width=.48\linewidth]{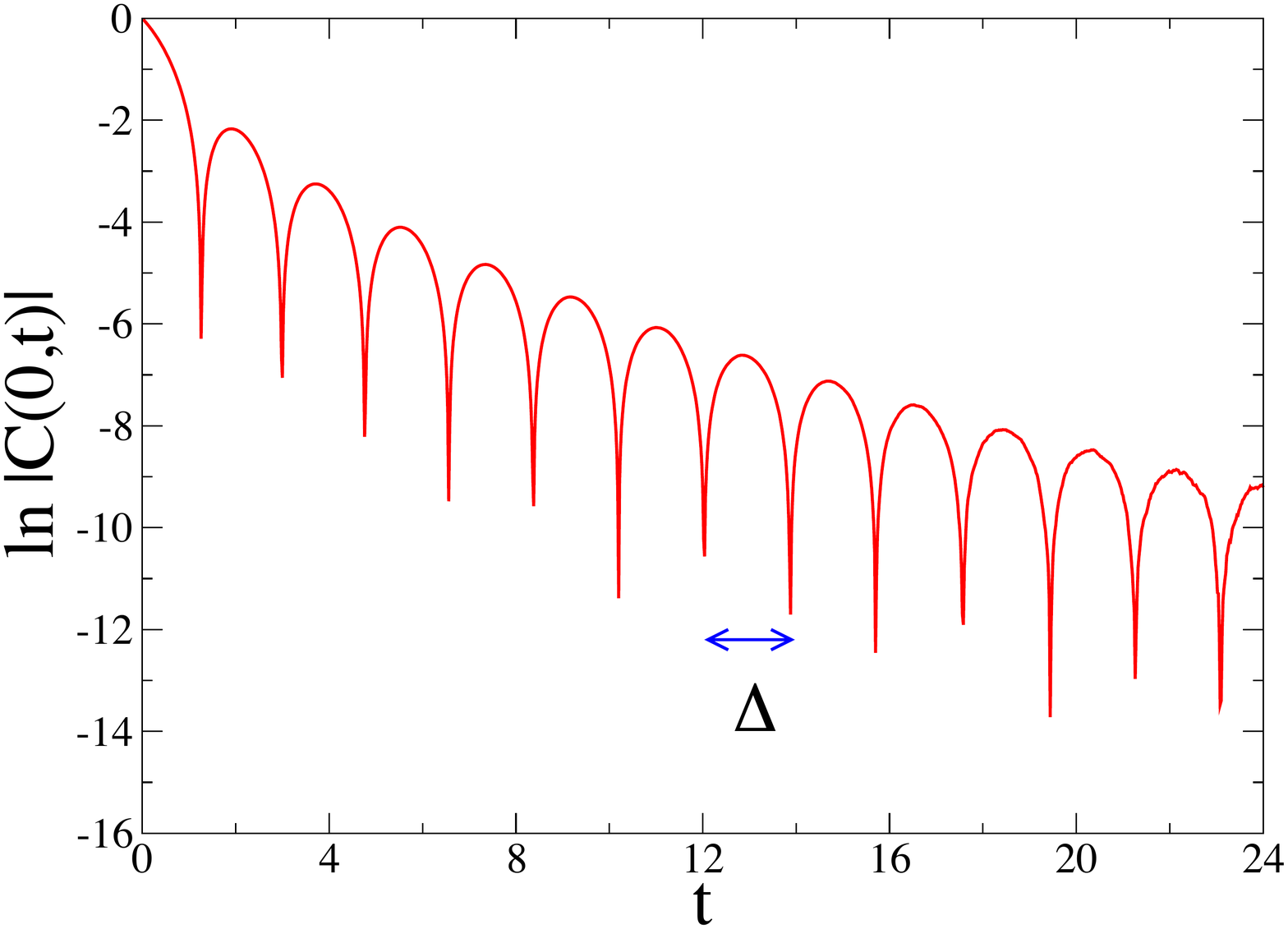}
{\hskip 4pt}
\includegraphics[angle=0,width=.48\linewidth]{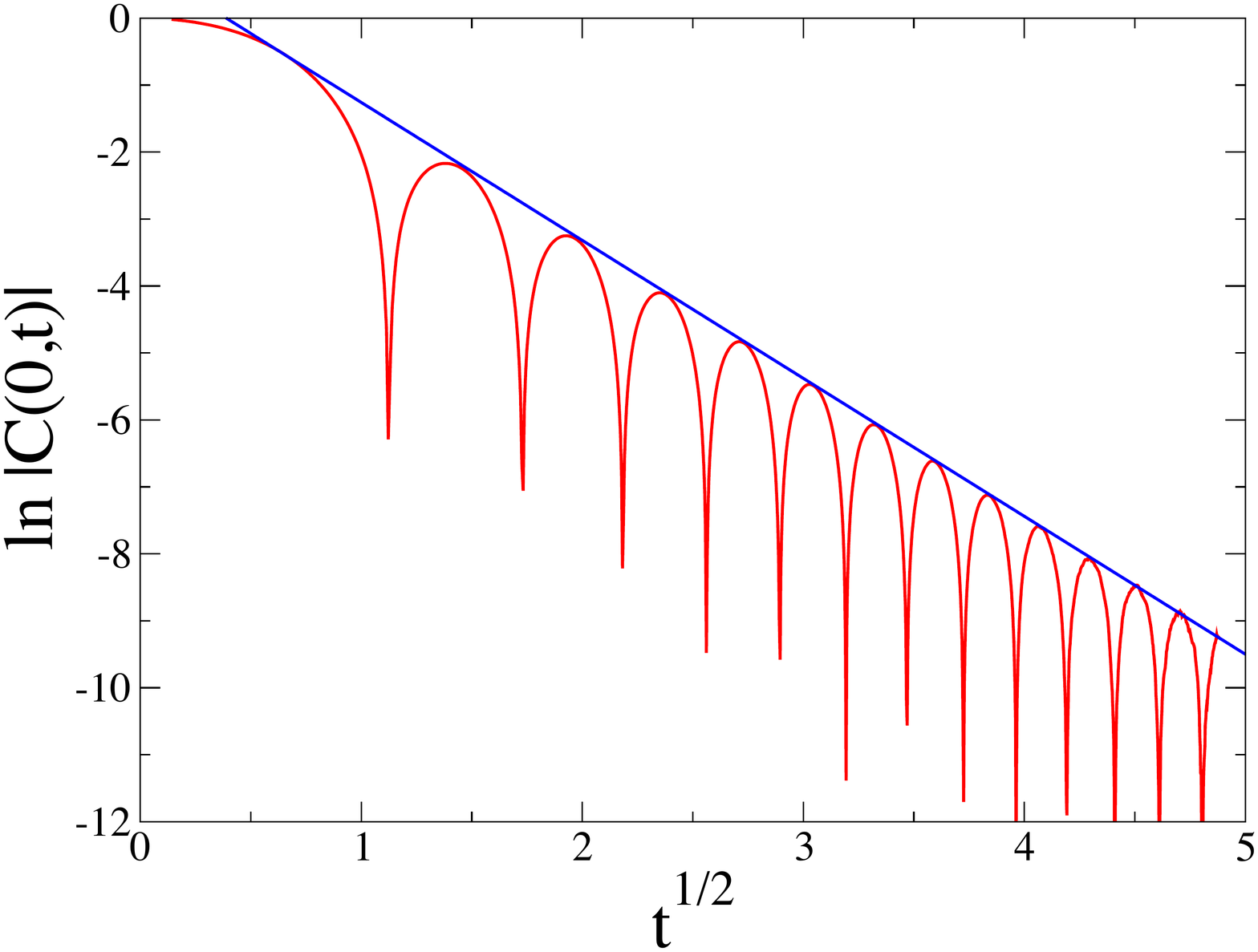}
\caption{\small
Absolute logarithmic plots of the correlation $C(0,t)$ at the TASEP point.
Left: plot against time $t$.
Bar with left-right arrows: half-period $\Delta\approx1.84$.
Right: plot against $\sqrt{t}$.
The straight line tangent to the data has an absolute slope $A\approx2.06$.}
\label{cc}
\end{center}
\end{figure}

Let us first present our numerical data for the TASEP point ($\eps=1$).
Figure~\ref{cc} shows absolute logarithmic plots of $C(0,t)$
against time $t$ (left) and against $\sqrt{t}$ (right).
The left panel shows the periodic pattern of oscillations,
with a first zero at $t_1\approx1.26$ and a half-period $\Delta\approx1.84$.
The slope of the straight line drawn on the right panel yields $A\approx2.06$
for $\eps=1$.

Repeating the same analysis for several positive values of $\eps$ along the ASEP line,
we obtain the data shown in figure~\ref{asep}.
The left panel shows the amplitude~$A$ of the stretched exponential
law~(\ref{caeps}) against $\eps$.
This quantity is observed to depart continuously from
the analytically known SEP value (see~(\ref{amax})), shown as a blue square.
The right panel shows the products $\eps t_1$ and $\eps\Delta$ against~$\eps$,
where~$t_1$ is the location of the first zero of $C(0,t)$,
while the semi-period~$\Delta$ is the asymptotic lapse of time
between two consecutive zeros.
Both products hardly depend on $\eps$.
So, the scaling laws
\beq
t_1\sim\Delta\sim\frac{1}{\eps},
\label{t1del}
\eeq
which hold exactly along the solvability line (see~(\ref{t1eps}),~(\ref{deleps})),
also hold approxima\-tely along the ASEP line.

\begin{figure}[!ht]
\begin{center}
\includegraphics[angle=0,width=.48\linewidth]{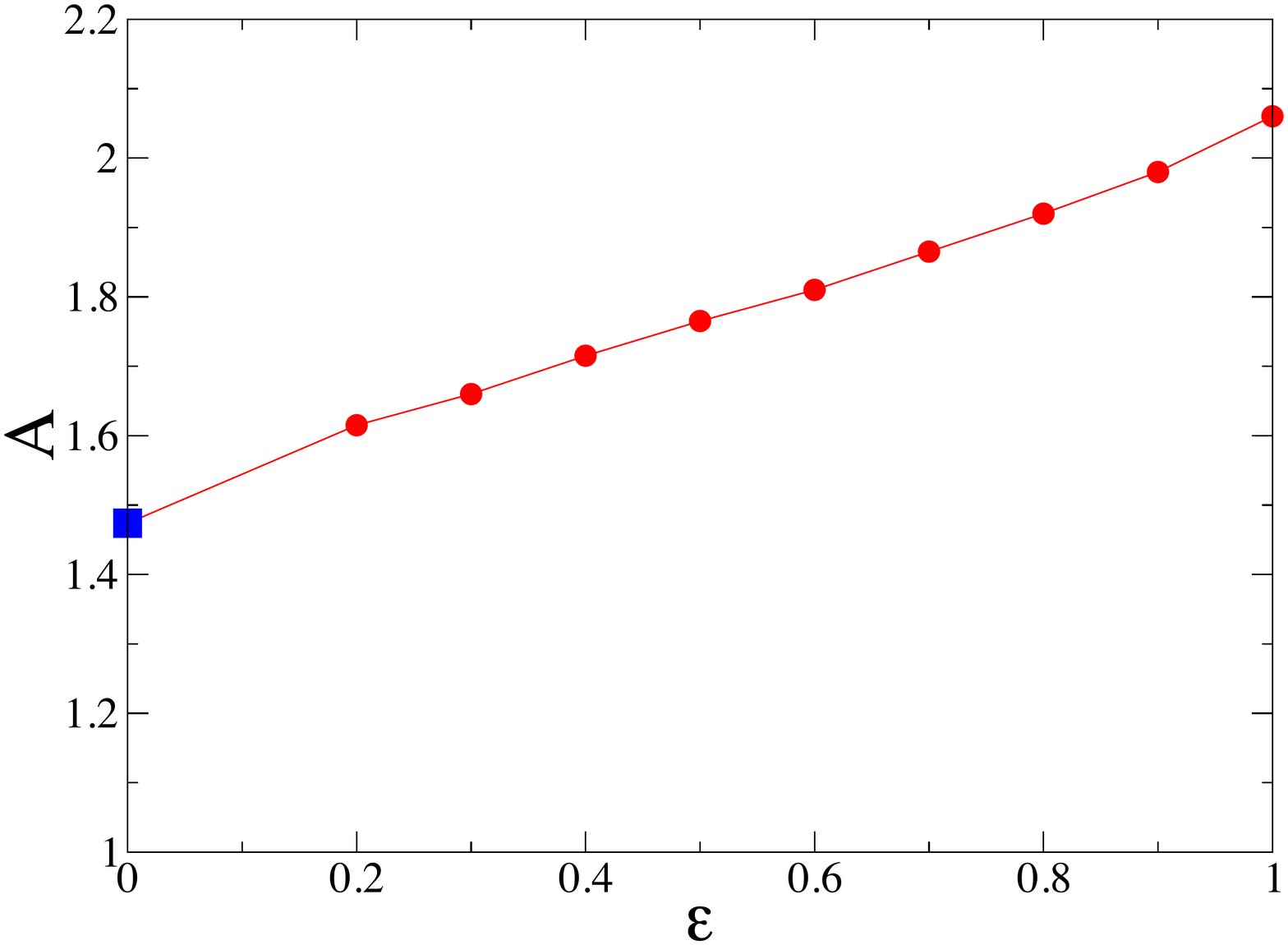}
{\hskip 4pt}
\includegraphics[angle=0,width=.48\linewidth]{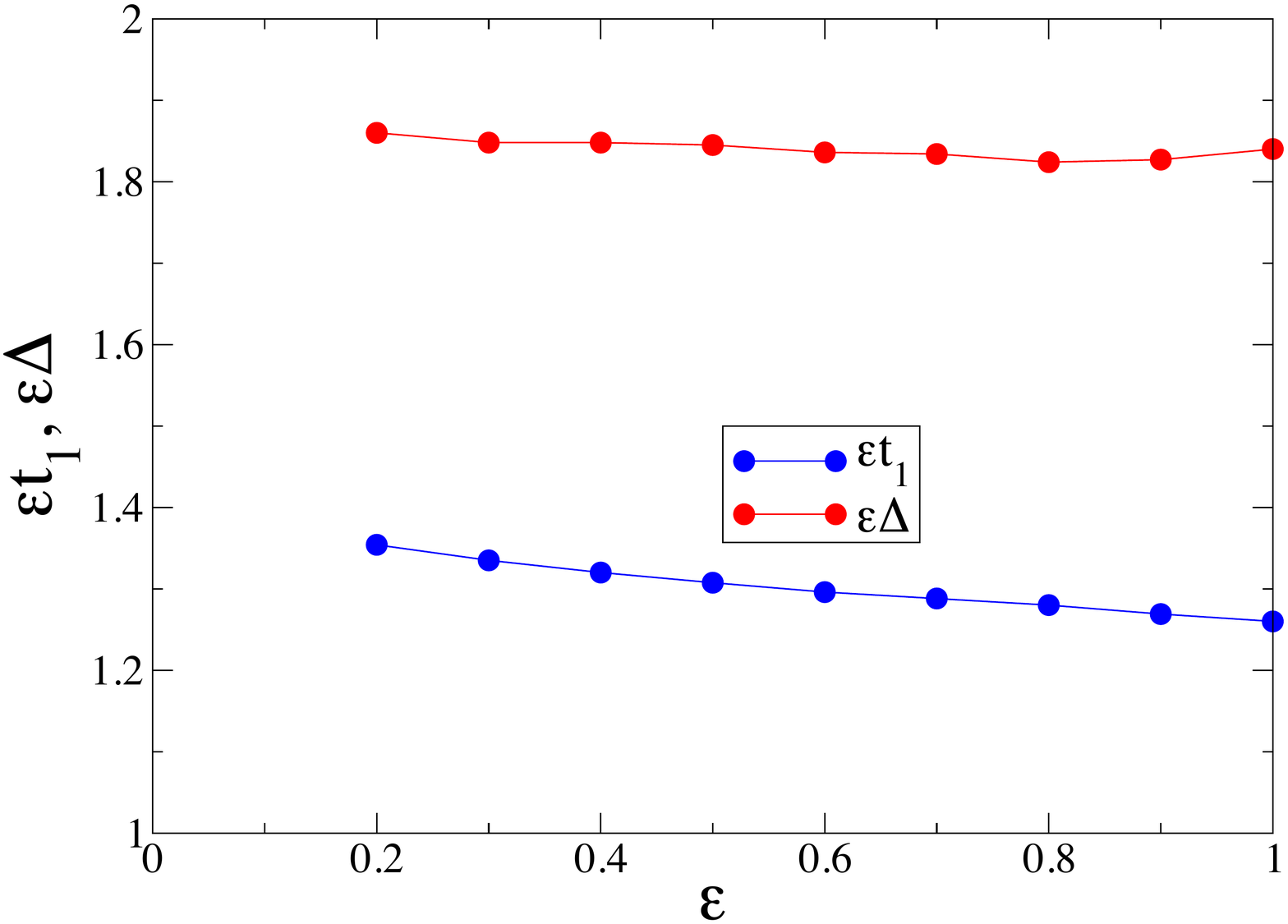}
\caption{\small
Left: amplitude $A$ of the stretched exponential law~(\ref{caeps})
along the ASEP line, against $\eps$.
Blue square: analytically known SEP value at $\eps=0$ (see~(\ref{amax})).
Right: products $\eps t_1$ and $\eps\Delta$
along the ASEP line, against~$\eps$.}
\label{asep}
\end{center}
\end{figure}

\subsection{Generic behaviour}
\label{generic}

We now turn to a brief description of the infinite-temperature dynamics
for generic values of the parameters $\delta$ and $\eps$ inside the triangular
region
shown in figures~\ref{fig:triangle} and~\ref{fig:triangle2}.
The correlation $C(0,t)$ falls off exponentially, as
\beq
C(0,t)\sim\e^{-\alpha_1 t},
\eeq
where the decay rate $\alpha_1$ has a rather weak dependence
on the irreversibility parameter~$\eps$.
This exponential decay is modulated
by oscillations which are asymptotically periodic in time.
Here again, both the first zero $t_1$ and the half-period~$\Delta$
roughly follow the $1/\eps$ law~(\ref{t1del}).

\begin{figure}[!ht]
\begin{center}
\includegraphics[angle=0,width=.48\linewidth]{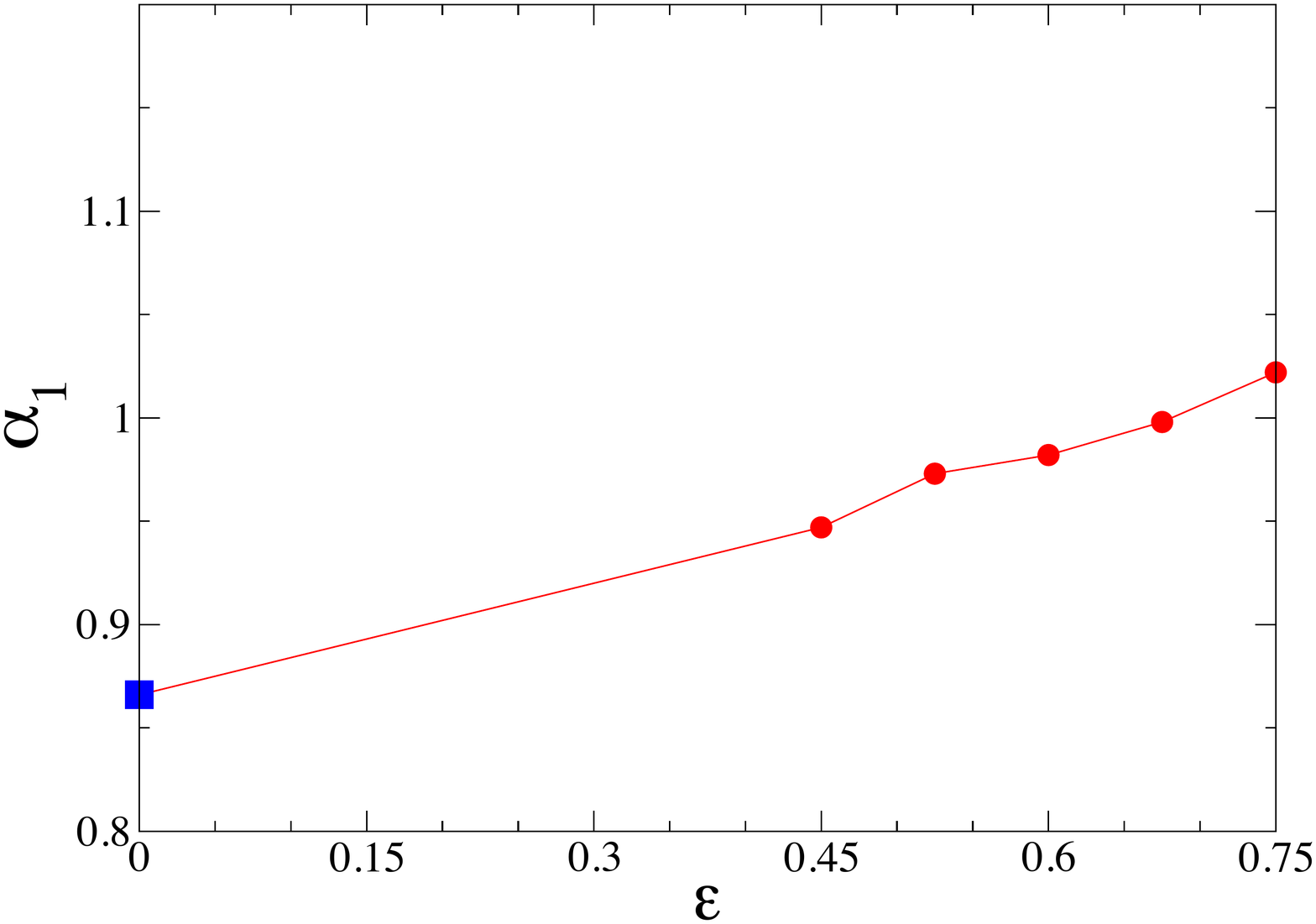}
{\hskip 4pt}
\includegraphics[angle=0,width=.48\linewidth]{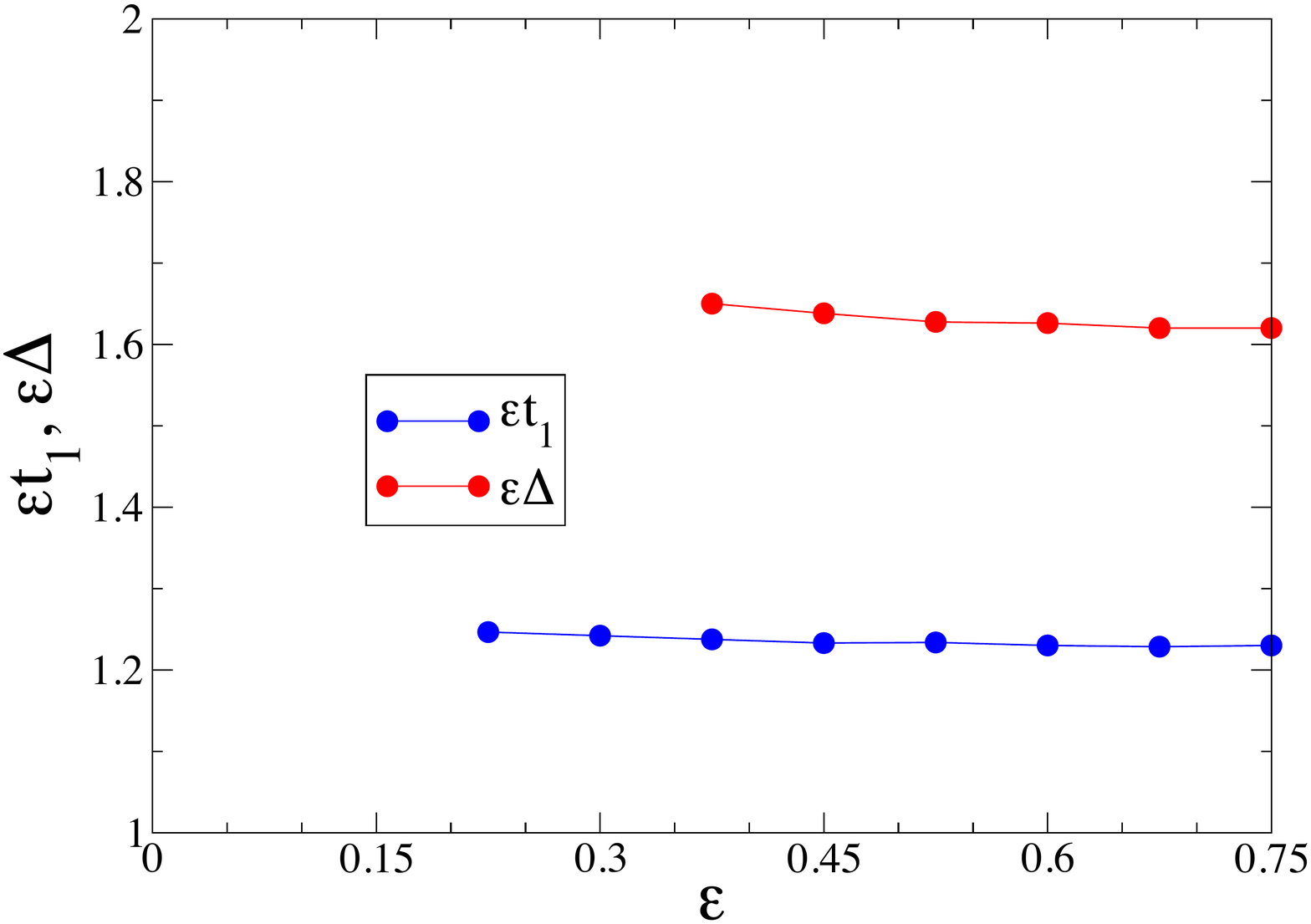}
\caption{\small
Left: decay rate $\alpha_1$ against $\eps$ for $\delta=-1/2$.
Blue square: known value $\alpha_1=\sqrt{3}/2$ at $\eps=0$ (see~(\ref{gap})).
Right: products $\eps t_1$ and $\eps\Delta$ against~$\eps$ for $\delta=-1/2$.}
\label{demi}
\end{center}
\end{figure}

These features are illustrated in figure~\ref{demi} for $\delta=-1/2$,
hence $\eps_\max=3/4$.
Data are only given whenever the plotted quantities can be measured
with a reasonable enough accuracy,
i.e., for not too small values of the irreversibility parameter $\eps$.
Even so, statistical errors are more important than along the ASEP line
(compare figures~\ref{asep} and~\ref{demi}).
The left panel shows the decay rate $\alpha_1$ against $\eps$.
This quantity seems to depart continuously
from the analytically known value $\alpha_1=\sqrt{3}/2$
at $\eps=0$ (see~(\ref{gap})), shown as a blue square.
The right panel shows the products $\eps t_1$ and $\eps\Delta$ against~$\eps$.
These products again exhibit a very weak dependence on $\eps$,
so the scaling laws~(\ref{t1del}) hold approximately
for generic values of the parameters $\delta$ and $\eps$.

\subsection{Spectrum of the Markov matrix}
\label{spectrum}

A useful alternative way of investigating the problem
consists in looking at the spectrum of the Markov matrix
which represents the generator of the stochastic dynamics on a finite chain.
This approach has already proved to be a useful tool
in several circumstances~\cite{melin,gm,slm}.
Here again, the shape of the spectra of Markov matrices
will provide a clear picture of the main characteristics of the dynamics,
including reversibility or integrability.

A finite system of $N$ sites with periodic boundary conditions
has $2^N$ configurations $\{\s_1\dots\s_N\}$,
that we assume to be ordered lexicographically from $\{+++\cdots\}$ to $\{---\cdots\}$.
Let $\vecP(t)$ denote the vector of the probabilities of these configurations at time $t$.
This vector obeys the differential equation
\beq
\frac{{\rm d}\vecP}{{\rm d}t}=\vecM\vecP,
\eeq
where the $2^N\times2^N$ matrix $\vecM$ is the Markov matrix of the problem.
Its non-diagonal entries are the transition rates between adjacent configurations
(i.e., configurations which differ by a single spin flip),
while its diagonal entries are determined by the rule that column sums vanish,
ensuring the conservation of probability.

If the dynamics is reversible ($\eps=0$, i.e., $w_{+-}=w_{-+}$),
the Markov matrix~$\vecM$ is symmetric and its eigenvalues are real.
Otherwise $\vecM$ has complex spectrum in general.
Let us henceforth denote the eigenvalues of $\vecM$ as $-\E_i$, with $\Re\E_i\ge0$.
As the dynamics obeys spin reversal symmetry,
the eigenvectors of $\vecM$ have a definite parity,
i.e., they are either even or odd under spin reversal.
Even and odd subspaces have equal dimensions $2^{N-1}$.
The eigenvalue $\E_1=0$, corresponding to the stationary state,
pertains to the even sector.
The stationary state is unique, so $\E_1$ is non-degenerate,
except in the extremal cases ($\delta=\pm1$) which obey conservation laws.
Finally, the spectra of the Markov matrix $\vecM$
and of the corresponding quantum Hamiltonian $\vecH$
(see section~\ref{sec:quantum}) are related as follows:
the spectrum of $\vecH$ consists of two copies
of the spectrum of $\vecM$ in the even sector,
while the eigenvalues of $\vecM$ in the odd sector do not appear in $\vecH$.

The smallest generic system consists of $N=3$ sites.
The corresponding $8\times8$ Markov matrix $\vecM$ and its eigenvalues
can be written down explicitly.
We have
\beq
\fl
\vecM=\pmatrix{
-3w_{++}&w_{++}&w_{++}&0&w_{++}&0&0&0\cr
w_{++}&-s&0&w_{-+}&0&w_{+-}&0&0\cr
w_{++}&0&-s&w_{+-}&0&0&w_{-+}&0\cr
0&w_{+-}&w_{-+}&-s&0&0&0&w_{++}\cr
w_{++}&0&0&0&-s&w_{-+}&w_{+-}&0\cr
0&w_{-+}&0&0&w_{+-}&-s&0&w_{++}\cr
0&0&w_{+-}&0&w_{-+}&0&-s&w_{++}\cr
0&0&0&w_{++}&0&w_{++}&w_{++}&-3w_{++}},
\eeq
where the rates $w_{++}$, $w_{+-}$, $w_{-+}$ are given in
table~\ref{tab:fmoves} in the infinite-temperature case ($\epsr=0$),
and with the shorthand $s=w_{++}+w_{+-}+w_{-+}$.
The eigenvalues of $\vecM$ are as follows.
\beqa
&&\mbox{Even sector:}\quad
\E_1=0,\quad
\E_2=2(1+\delta),\quad
\E_{3,4}=2-\delta\pm\ii\,\eps\sqrt3.
\nonumber\\
&&\mbox{Odd sector:}\quad
\ \E_{5,6}=2\pm\sqrt{1+3\delta^2},\quad
\E_{7,8}=1\pm\ii\,\eps\sqrt3.
\eeqa
These expressions are not invariant under the change $\delta\to-\delta$.
This symmetry of the infinite chain indeed only holds
on systems whose size $N$ is a multiple of 4.

Let us now describe the spectrum of the Markov matrix
of infinite-temperature dynamics on larger systems.
The following discussion (see the contrast between figures~\ref{int},~\ref{sreversible}
and~\ref{clouds}) clearly demonstrates that an investigation of these spectra
provides a very sensitive tool
to detect characteristic features of the underlying dynamics,
such as integrability or irreversibility.

\begin{itemize}

\item
{\it Infinite-temperature Glauber model ($\delta=\eps=0$).}
As already mentioned, this case corresponds to a dynamics of independent spins.
The result~(\ref{idt}) generalizes as follows.
The correlation function built on any $k$uple of distinct spins decays exponentially as
\beq
\mean{\s_{i_1}(t)\dots\s_{i_k}(t)\s_{i_1}(0)\dots\s_{i_k}(0)}=\e^{-kt}.
\label{composite}
\eeq
The spectrum of the Markov matrix is therefore highly degenerate~\cite{slm}:
it consists of the integers $k=0,\dots,N$, with the combinatorial multiplicities
\beq
\mu(N,k)=\pmatrix{N\cr k}=\frac{N!}{k!(N-k)!}.
\eeq
The even (resp.~odd) sector corresponds to even (resp.~odd) values of~$k$.

\item
{\it Solvability line ($\delta=0$).}
Along this line, parametrized by the irreversibility parameter $\eps$,
the eigenvalues of the Markov matrix are still given by simple formulas.
They are indeed observed to be linear combinations,
with integer coefficients, of the following complex frequencies
\beq
\Omega_n=1-\ii\l_n,\quad\l_n=2\eps\sin\frac{n\pi}{N}.
\label{wn}
\eeq
The latter frequencies are obtained by inserting into the dispersion law~(\ref{Omegaq})
discrete quantized momenta of the form $q=n\pi/N$,
corresponding to both periodic and anti-periodic boundary conditions.
So, the real part of any eigenvalue is still an integer $k=0,\dots,N$,
while its imaginary part is an integer linear combination of the $\l_n$.
It is therefore strictly proportional to $\eps$.
Figure~\ref{int} shows the upper parts ($\Im\E\ge0$) of the spectra
of the Markov matrices for $N=6$ and $N=7$, with $\delta=0$ and $\eps=\eps_\max=1/2$.
The imaginary parts of the eigenvalues, shown by horizontal lines,
read in increasing order (bottom to top):
For $N=6$ (left):
$\l_1$, $\l_2$, $\l_3$, $\l_1+\l_3$, $2\l_2$.
For $N=7$ (right):
$\l_3-\l_2$, $\l_1+\l_2-\l_3$, $\l_2-\l_1$, $\l_1$, $\l_3-\l_1$,
$\l_1+\l_3-\l_2$,
$\l_2$, $\l_3$, $\l_1+\l_2$, $\l_2+\l_3-\l_1$, $\l_1+\l_3$, $\l_2+\l_3$,
$\l_1+\l_2+\l_3$.

\begin{figure}[!ht]
\begin{center}
\includegraphics[angle=0,width=.48\linewidth]{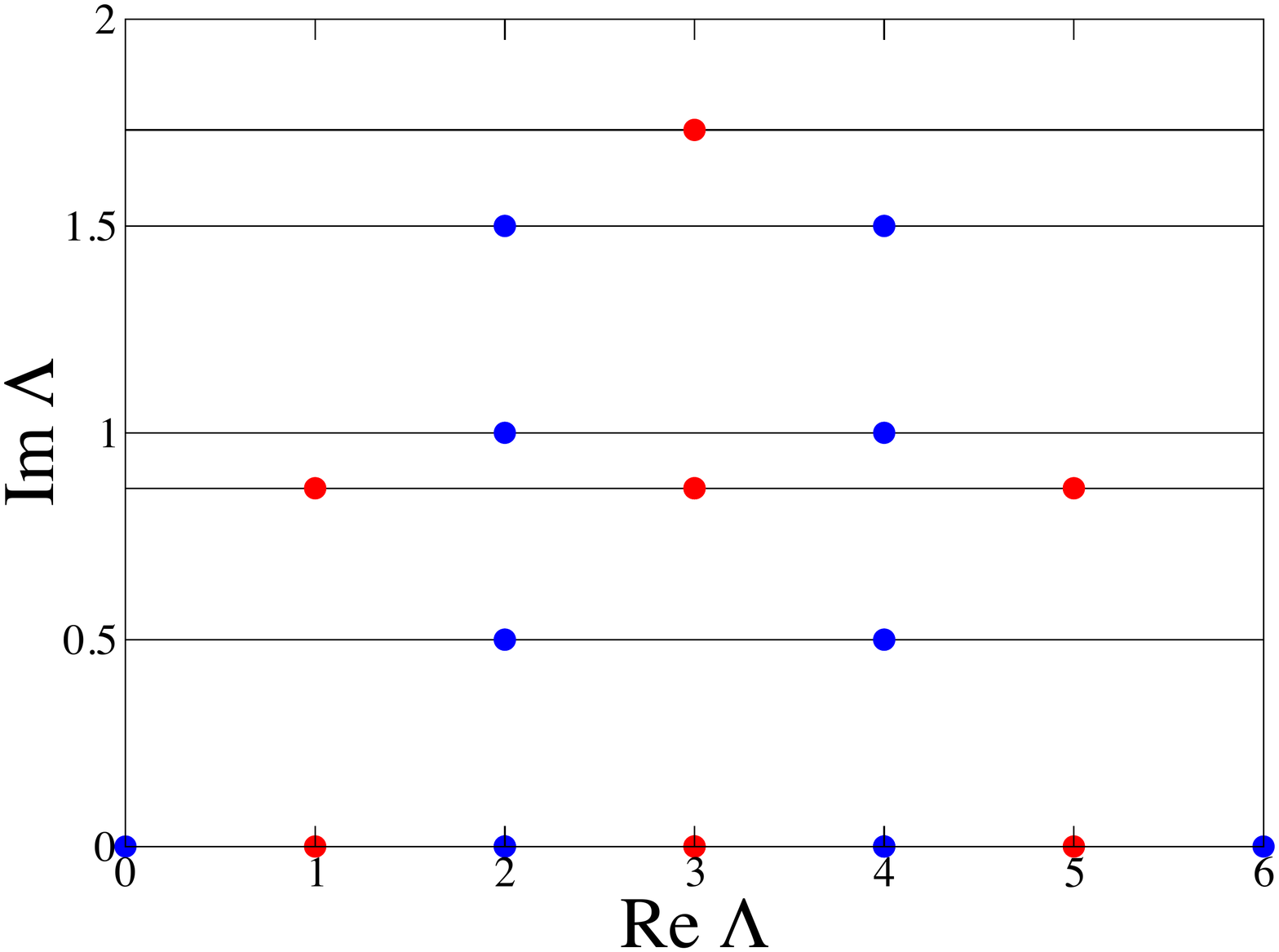}
{\hskip 4pt}
\includegraphics[angle=0,width=.48\linewidth]{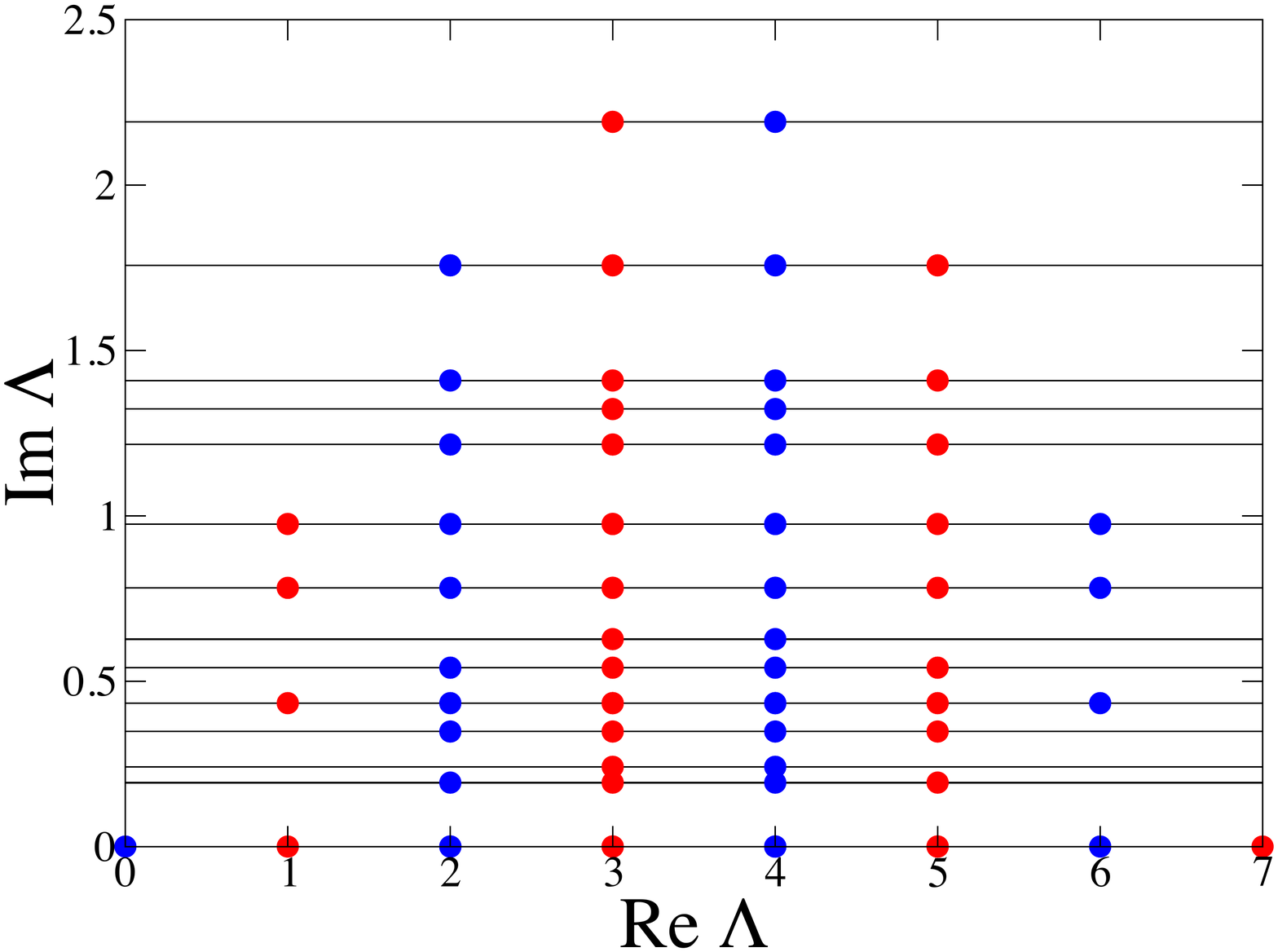}
\caption{\small
Upper parts ($\Im\E\ge0$) of the spectra of the infinite-temperature Markov
matrices $\vecM$ for $N=6$ (left) and $N=7$ (right)
with $\delta=0$ and $\eps=\eps_\max=1/2$.
Blue symbols: even sector.
Red symbols: odd sector.
Horizontal lines: imaginary parts listed in the text.}
\label{int}
\end{center}
\end{figure}

\item
{\it Reversibility line ($\eps=0$).}
Along this line, parametrized by the non-linearity parameter $\delta$,
the spectrum of the Markov matrix is real.
Figure~\ref{sreversible} shows this spectrum for $N=8$ against $\delta$.
The plot is invariant under the change $\delta\to-\delta$,
as expected as the system size is a multiple of 4.
The highly degenerate integer spectrum of the Glauber point
is manifest in the middle of the plot ($\delta=0$).
Level crossings take place in both sectors.
The lowest eigenvalue $\E=0$, corresponding to the stationary state,
is non-degenerate, except at the endpoints ($\delta=\pm1$),
where the conservation laws induce degeneracies.

\begin{figure}[!ht]
\begin{center}
\includegraphics[angle=0,width=.7\linewidth]{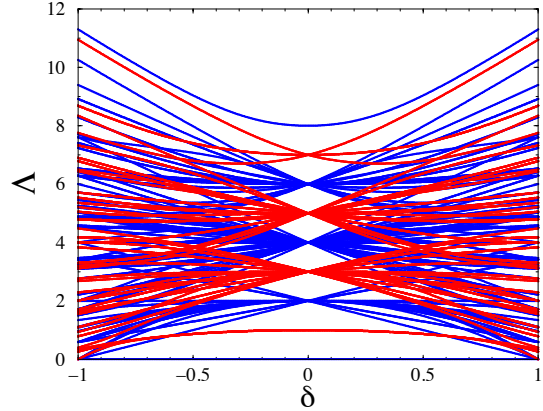}
\caption{\small
Spectrum of the Markov matrix $\vecM$ of the infinite-temperature reversible
dynamics
against the non-linearity parameter $\delta$ for $N=8$.
Blue lines: even sector.
Red lines: odd sector.}
\label{sreversible}
\end{center}
\end{figure}

\item
{\it Generic case.}
For arbitrary parameter values ($\delta\ne0$ and $\eps\ne0$),
the spectrum of the Markov matrix appears as a rather structureless cloud
in the complex $\E$-plane.
Figure~\ref{clouds} shows this spectrum for $N=12$ in two cases.
At the TASEP point ($\delta=-1$, $\eps=1$) (left),
eigenvalues tend to accumulate near $\Re\E=0$.
This observation goes hand in hand with the fact that the eigenvalue $\E=0$
is degenerate, because the dynamics conserves the total energy.
At a generic point ($\delta=-1/2$, $\eps=2/3$) (right),
there are fewer eigenvalues near $\Re\E=0$.
The outermost part of the spectrum shows alternating blue and red stripes,
which are remnants of the ordered structure of the spectrum in the solvable case.

\begin{figure}[!ht]
\begin{center}
\includegraphics[angle=0,width=.48\linewidth]{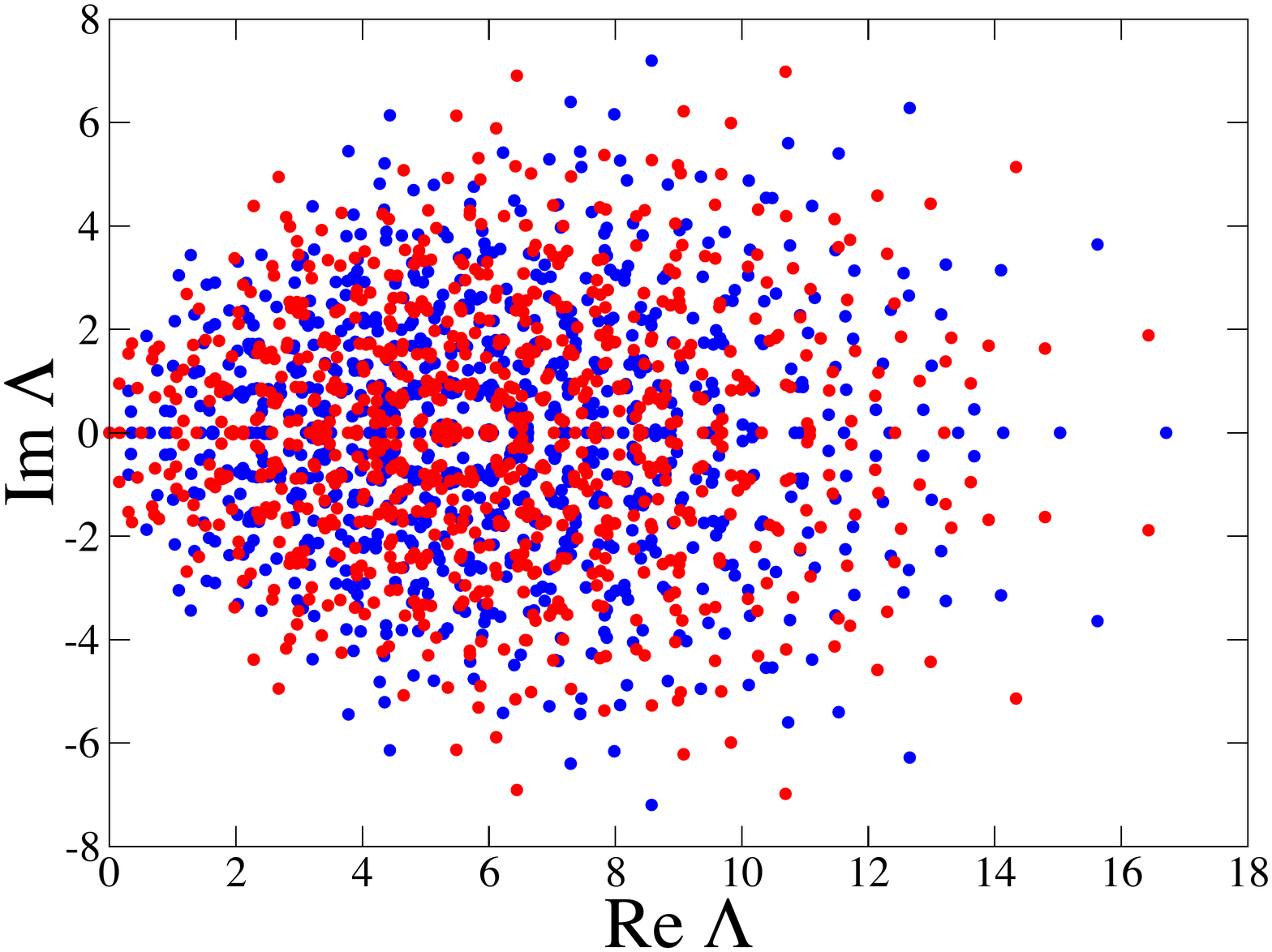}
{\hskip 4pt}
\includegraphics[angle=0,width=.48\linewidth]{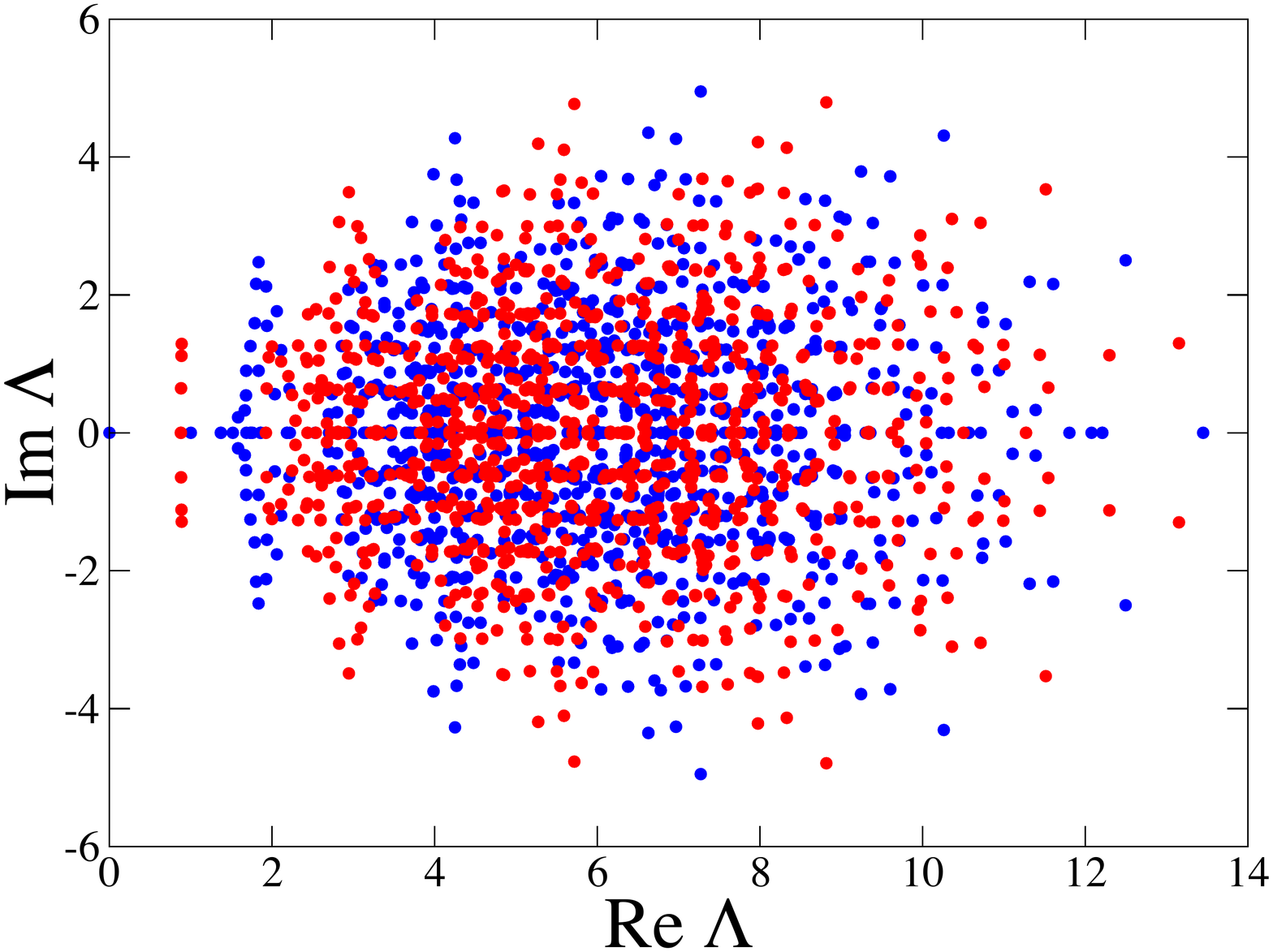}
\caption{\small
Spectrum in the complex $\E$-plane
of the infinite-temperature Markov matrix $\vecM$ for $N=12$.
Left: TASEP point.
Right: a generic point ($\delta=-1/2$, $\eps=2/3$).
Blue symbols: even sector.
Red symbols: odd sector.}
\label{clouds}
\end{center}
\end{figure}

\end{itemize}

\section{Finite temperature}
\label{finite}

In this section we focus our attention onto the novel features
of the finite-temperature dynamics which were absent in the infinite-temperature situation,
investigated in section~\ref{sec:infinite}.

\subsection{Reversibility line ($\eta=0$)}
\label{freversible}

The finite-temperature reversible dynamics exhibits several novel features
with respect to its infinite-temperature counterpart
(see section~\ref{reversibleline}).

The correlation functions $C(0,t)$ (random initial state)
and $C_\stat(t)$ (thermalized initial state) are different from each other.
The analysis of the solvable case ($\delta=0$) recalled in
section~\ref{sec:reminder} however strongly suggests that,
for a reversible dynamics, both correlations fall off exponentially
with a common decay rate $\alpha_1$, which depends both on temperature and
on the non-linearity parameter $\delta$.

The first few coefficients of the time series expansion~(\ref{cseries})
of the correlation function $C(0,t)$ read
\beqa
a_0&=&1,
\nonumber\\
a_1&=&1,
\nonumber\\
a_2&=&1+2\epsr^2+\delta^2,
\nonumber\\
a_3&=&1+6\epsr^2+8\epsr^2\delta+5\delta^2,
\nonumber\\
a_4&=&1+12\epsr^2+6\epsr^4+48\epsr^2\delta+2(9+16\epsr^2)\delta^2+3\delta^4.
\label{fak}
\eeqa
The invariance of the infinite-temperature reversible dynamics
under the change $\delta\to-\delta$ is broken at any finite temperature.
This is testified by the presence of a term proportional to $\epsr^2\delta$ in
the coefficient $a_3$ in~(\ref{fak}).
Therefore, at variance with the infinite-temperature situation (see~(\ref{gap})),
the decay rate $\alpha_1$ is not expected to be even in $\delta$.
In particular the dynamics at the two endpoints $(\delta=\pm1$) are now different.
For $\delta=-1$, we recover the SEP point and its microcanonical dynamics,
investigated in section~\ref{sec:sep}, irrespective of temperature.
For $\delta=1$,
the allowed moves are the creation and annihilation of pairs of domain walls,
with respective rates $1\mp\g$.
This temperature-dependent dynamics conserves the total number of dual particles
described by the occupation numbers $\w\tau_n$ (see~(\ref{wtau})).
In the zero-temperature limit, this non-generic dynamics becomes
an interesting example of a kinetically constrained model,
which exhibits all the generic features of metastability.
This model has been investigated in detail in~\cite{gds},
together with various other one-dimensional examples.

The corresponding quantum Hamiltonian reads
\beq
\fl
\vecH=\half\sum_n\biggl(1
+\frac{v^2\delta-1}{1+v^2}\,\s_n^x\s_{n+1}^x
+\frac{\delta-v^2}{1+v^2}\,\s_n^y\s_{n+1}^y
+\delta\s_n^z\s_{n+1}^z+\frac{2v(1+\delta)}{1+v^2}\,\s_n^z\biggr),
\label{fham}
\eeq
where we recall that $v$ is a shorthand notation for $\tanh\beta$.
This Hamiltonian was already derived by N\'emeth~\cite{nemeth}
(up to a global factor 2 and with a different convention for the sign of $\s_n^z$).
It describes the XYZ (fully anisotropic Heisenberg) spin chain in a uniform external field.
This model is known not to be integrable in general.
As a consequence,
we have no analytical prediction for the relaxation rate $\alpha_1$,
except in the solvable case ($\delta=0$), where $\alpha_1=1-\g$
(see~(\ref{fgap})), and at the endpoints ($\delta=\pm1$),
where ferromagnetic interactions become isotropic,
so that $\vecH$ becomes invariant under spin rotations around the $z$ axis.
This symmetry goes hand in hand with the conservation laws.
As already underlined in~\cite{nemeth},
it implies the vanishing of $\alpha_1$ at these points.
Furthermore, the relaxation rate $\alpha_1$ can also be determined exactly
at a special point known as the KDH point (see section~\ref{sec:kdh}).
Finally, yet other special cases of the Hamiltonian~(\ref{fham})
are integrable as they correspond to physical realizations of Hecke algebras~\cite{ar}.
The latter cases however do not bring more results on the present problem.

\begin{figure}[!ht]
\begin{center}
\includegraphics[angle=0,width=.7\linewidth]{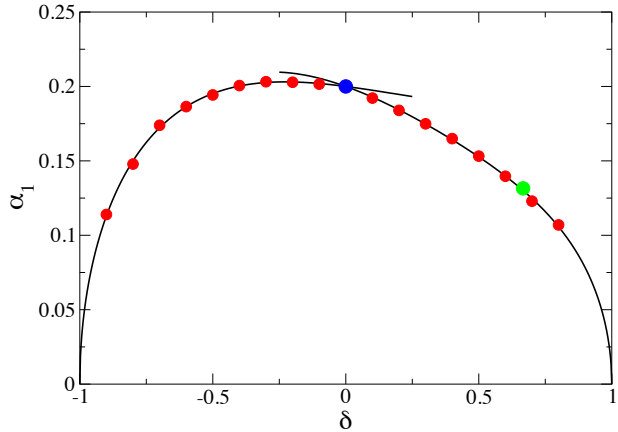}
\caption{\small
Relaxation rate $\alpha_1$ of the correlations $C(0,t)$ and $C_\stat(t)$
on the finite-temperature reversibility line ($\eta=0$)
against the non-linearity parameter $\delta$ for $v=1/2$, i.e., $\g=4/5$.
Blue symbol: exact result $\alpha_1=1-\g=1/5$ in the solvable case
($\delta=0$).
Green symbol: exact result for the KDH point (see~(\ref{aid})).
Red symbols: numerical data.
Full curves: fits (see text).}
\label{reversible}
\end{center}
\end{figure}

In order to investigate the dependence of the relaxation rate $\alpha_1$
on the non-linearity parameter $\delta$, we have measured the correlation $C(0,t)$ by means of numerical simulations.
Figure~\ref{reversible} shows our results
for a case of moderate temperature ($v=1/2$, i.e., $\g=4/5$).
The plotted values of $\alpha_1$ are reasonably accurate.
The estimated error bar is comparable to the symbol size on the figure.
Furthermore, we have checked that the stationary correlation $C_\stat(t)$
yields compatible values of $\alpha_1$ within the error bar.
The extrapolation procedure looses its accuracy near the endpoints ($\delta\to\pm1$),
where~$\alpha_1$ is known to vanish.
This is to be expected, as the correlation exhibits a crossover
to the stretched exponential law~(\ref{carho}) as $\delta\to-1$,
and to a similar kind of subexponential relaxation law as $\delta\to1$.
The plotted values for $\alpha_1$ seem to vanish according to the same
square-root law as the infinite-temperature result~(\ref{gap}).
They are also compatible with the presence
of a slight cusp at the solvable point ($\delta=0$),
shown as a blue symbol, where $\alpha_1=1-\g=1/5$.
The full curves show the outcomes of two separate fits
of the data points for $\delta\le0$ and $\delta\ge0$,
where $\alpha_1$ is assumed to be the product
of the result~(\ref{gap}) by a second-degree polynomial.

\subsection{Generic behaviour}
\label{fgeneric}

In the generic situation of an irreversible dynamics,
the most natural question is whether the non-trivial
phase diagram in the $\delta$--$\eta$ plane depicted in figure~\ref{regimes}
subsists at finite temperature.

We have evidenced, by means of numerical simulations,
the existence of the threshold $\eta_0$ between a regime
where the correlation $C(0,t)$ falls off monotonically
and a regime where it is oscillatory.
We have not addressed the more delicate question
(from a mere numerical standpoint)
of the existence of the other threshold~$\eta_c$, related to the non-analyticity
of the decay rate of the thermalized correlation $C_\stat(t)$.

\begin{figure}[!ht]
\begin{center}
\includegraphics[angle=0,width=.7\linewidth]{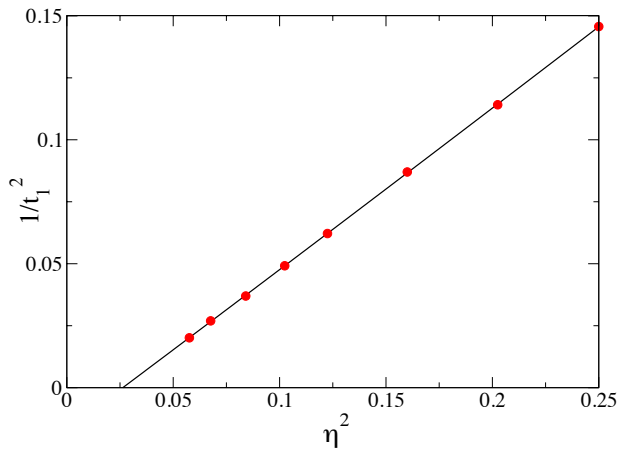}
\caption{\small
Plot of $1/t_1^2$ against $\eta^2$ for $\g=1/2$ and $\delta=-0.7$.
Red symbols: data points obtained by numerical simulation.
Full curve: quadratic fit yielding the threshold $\eta_0=0.162$.}
\label{t1plot}
\end{center}
\end{figure}

The threshold $\eta_0$ between the oscillatory and monotonic regimes
has been determined as the value of $\eta$ where $t_1$,
the first zero of the correlation $C(0,t)$, diverges.
Figure~\ref{t1plot} shows $1/t_1^2$ against $\eta^2$ for $\g=1/2$ and $\delta=-0.7$.
This way of analyzing data is inspired from the analytical
result~(\ref{ft1eta}) in the solvable case ($\delta=0$).
The data points (red symbols) exhibit an almost perfect linear law.
A quadratic fit (full curve) yields the threshold $\eta_0=0.162$.

\begin{figure}[!ht]
\begin{center}
\includegraphics[angle=0,width=.7\linewidth]{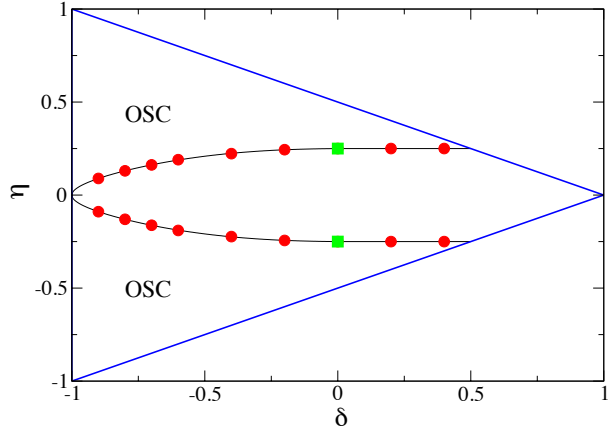}
\caption{\small
Phase diagram in the $\delta$--$\eta$ plane for $\g=1/2$.
Red symbols: data for the threshold $\pm\eta_0$.
Full black curves: fits suggesting the square-root law~(\ref{seuilsq}).
Green squares: exact values $\pm\eta_0=\pm 1/4$ for $\delta=0$ (see
figures~\ref{fig:triangle} and~\ref{fig:triangle2}).}
\label{seuil}
\end{center}
\end{figure}

Repeating the same analysis for several values of $\delta$,
we obtain the phase diagram in the $\delta$--$\eta$ plane
shown in figure~\ref{seuil}, for $\g=1/2$.
Symbols show the dependence on~$\delta$ of our prediction for the threshold
$\eta_0$ (and its symmetric counterpart $-\eta_0$).
The oscillatory regions ($\abs{\eta}>\eta_0$) are marked OSC.
The threshold $\eta_0$ is observed to be nearly constant for $\delta\ge0$
and equal to its value $\eta_0=\g/2=1/4$ in the solvable case ($\delta=0$)
(see~(\ref{eq:thresh})), shown as green squares.
The fit (full curve) suggests that the threshold vanishes with a square-root singularity,
\beq
\eta_0\sim\sqrt{1+\delta},
\label{seuilsq}
\eeq
as the SEP point is approached ($\delta\to-1$).

To close, let us underline that the existence of the threshold $\eta_0$
in the $\delta$--$\eta$ plane is not visible on the spectra of Markov matrices,
which always look like the infinite-temperature spectra shown in figure~\ref{clouds}.

\subsection{Equal-time spin-correlation function}
\label{etsc}

Another novel feature of finite-temperature dynamics
is the non-triviality of the equal-time spin-correlation function $C_n(t)$
(see~(\ref{eqts})), starting from a random initial state.

Here we focus our attention onto the correlation between neighboring spins
$C_1(t)=\mean{\s_0(t)\s_1(t)}$, i.e., (minus) the ferromagnetic energy density
per site (or bond).
At infinite temperature, for an arbitrary spatially homogeneous initial state,
this quantity obeys the closed evolution equation
\beq
\frac{{\rm d}C_1(t)}{{\rm d}t}=-2(1+\delta)C_1(t).
\label{einf}
\eeq
For a random initial condition ($C_1(0)=0$),
we thus recover that $C_1(t)$ remains equal to zero all throughout its evolution.

At finite temperature, in contrast, $C_1(t)$ increases monotonically from zero
towards its equilibrium value $C_1=\tanh \beta$ (see~(\ref{cstat})).
Its time series expansion reads
\beq
C_1(t)=\sum_{k\ge1}b_k(-1)^{k-1}\frac{t^k}{k!},
\label{cseries2}
\eeq
with
\beqa
b_1&=&\g(1+\delta),
\nonumber\\
b_2&=&2\g(1+\delta)^2,
\nonumber\\
b_3&=&\g(1+\delta)^3(4+\g^2),
\nonumber\\
b_4&=&2\g(1+\delta)^3(4(1+\g^2)(1+\delta)-\g^2),\ \dots
\label{eq:expansion}
\eeqa

It is striking to observe that there is no dependence in the irreversibility
parameter~$\eta$ before the 7th order.
At this order, $b_7$ contains a term of the form
$8\g^3(1+\delta)^3\delta^2\eta^2$.
Similarly, $b_8$ contains a term of the form
$24\g^3(1+\delta)^3(8+\delta)\delta^2\eta^2$,
while $b_9$ contains one term in $\eta^2$ and one term in $\eta^4$.
As it turns out, $\eta^2$ is always accompanied by $\delta^2$,
as the correlation $C_n(t)$ does not depend on the irreversibility
parameter $\eta$ at all in the solvable case ($\delta=0$) (see section~\ref{sec:linthe}).
A similar very weak effect of the irreversibility on equal-time spin correlations
has been observed both numerically and by short-time expansions
for the two-dimensional Ising model in its high-temperature phase~\cite{pleim2014}.

At high temperature, the full temporal behaviour of $C_1(t)$ can be derived as follows.
To first order in $\g$, the coefficients of the time series expansion~(\ref{cseries2})
assume the simple form $b_k\approx2^{k-1}(1+\delta)^k\g$, irrespective of $\eta$.
This has been checked up to the 9th order.
Hence
\beq
C_1(t)\approx\frac{\g}{2}\left(1-\e^{-2(1+\delta)t}\right).
\eeq The asymptotic value $C_1\approx\g/2$ agrees with the exact result, i.e., $v$,
as $\g\approx2v\approx2\beta$ to leading order at high temperature.
The above result suggests that the convergence rate $\alpha_E$ of the energy,
such that
\beq
\tanh\beta-C_1(t)\sim\e^{-\alpha_E t},
\eeq
takes the simple value
\beq
\alpha_E=2(1+\delta)
\label{aeres}
\eeq
to leading order in the high-temperature regime.
This result coincides with the prefactor in the right-hand side of~(\ref{einf}).
It vanishes when $\delta=-1$, as should be, since this corresponds to the SEP point
whose microcanonical dynamics conserves the total energy.
Finally, the rate $\alpha_E$ is also mentioned in~\cite{jb}
as being one of the gaps governing the low-temperature thermodynamics
of the quantum XXZ model.

At finite temperature, we have no analytical prediction for $\alpha_E$ in general,
which is expected to have a (weak) dependence on the irreversibility parameter $\eta$.
In the solvable case ($\delta=0$),
we have $\alpha_E=2\alpha_1=2(1-\g)$, irrespective of $\eta$~\cite{cg11}.

\subsection{A special point on the reversibility line}
\label{sec:kdh}

The dynamics simplifies for the special point considered by Kimball~\cite{kdh1}
and by Deker and Haake~\cite{kdh2}, where
\beq
\delta=\frac{\g}{2-\g},
\eeq
yielding
\beq
\label{kdh}
w_n
=\frac{1}{2}\left(1-\frac{\gamma}{2-\gamma}(\s_n(\s_{n-1}+\s_{n+1})-\s_{n-1}\s_{n+1})\right).
\eeq
In the $\delta$--$\epsilon$ plane, this point is located at the intersection of
the reversibility line with the second bisectrix, and therefore obeys the
condition $\delta=-\epsr$.
The KDH point is alternatively characterised by the fact that three
of the four rates coincide: $w_{++}=w_{+-}=w_{-+}$, as can be seen on~(\ref{kdh}).

For an arbitrary reversible dynamics, and an arbitrary magnetized initial state,
the evolution equation for the magnetization profile $\mean{\s_n}$ reads
\beq\label{eq:mag}
\frac{\rm d}{{\rm d}t}\langle\s_n\rangle
=-\langle\s_n\rangle-\epsr(\langle\s_{n-1}\rangle+\langle\s_{n+1}\rangle)
-\delta\,\langle\pi_n\rangle,
\eeq
where
\beq
\pi_n=\s_{n-1}\s_n\s_{n+1}.
\eeq
The evolution equation of the latter quantity reads
\beq\label{eq:q}
\frac{\rm d}{{\rm d}t}\langle\pi_n\rangle
=-3\langle\pi_n\rangle
-2\epsr(\langle\s_{n-1}\rangle+\langle\s_{n+1}\rangle)
-\delta\,\langle\s_n\rangle+\langle\phi_n\rangle,
\eeq
where
\beq
\fl
\phi_n=-\epsr(\s_{n-2}\s_n\s_{n+1}+\s_{n-1}\s_n\s_{n+2})-\delta(\s_{n-1}\s_{n+1}\s_{n+2}+\s_{n-2}\s_{n-1}\s_{n+1}).
\eeq
Consider a system of $N$ sites with periodic boundary conditions.
If $\epsr=-\delta$, the spatial sum of the $\phi_n$ vanishes identically.
Defining
\beq
M(t)=\frac{1}{N}\sum_n\langle\s_n\rangle,\quad
\Pi(t)=\frac{1}{N}\sum_n\langle\pi_n\rangle,
\eeq
we obtain from~(\ref{eq:mag}) and (\ref{eq:q}) two coupled linear equations
for these quantities:
\beq
\frac{\rm d}{{\rm d}t}\pmatrix{M(t)\cr\Pi(t)}
=\pmatrix{2\delta-1&-\delta\cr 3\delta&-3}\pmatrix{M(t)\cr\Pi(t)}.
\eeq
The corresponding decay rates are the opposites of the eigenvalues
of the above matrix, i.e.~\cite{kdh1,kdh2,malte},
\beq
\alpha_\pm=2-\delta\pm\sqrt{1+2\delta-2\delta^2}
=\frac{4-3\g\pm\sqrt{4-3\g^2}}{2-\g}.
\eeq

For a ferromagnetic model, it is physically reasonable to assume
that the spin autocorrelation $C(0,t)$ falls off at the same rate
as the total magnetization $M(t)$.
This line of thought yields the identity
\beq
\alpha_1=\alpha_-.
\label{aid}
\eeq
For $\g=4/5$,
the KDH point is at $\delta=2/3$ and~(\ref{aid}) reads
$\alpha_1=(4-\sqrt{13})/3=0.131482$.
This prediction, shown in figure~\ref{reversible} as a green symbol,
agrees with our numerical results.

It has been noticed in~\cite{malte}
that a similar construction works for the reversible dynamics with
\beq
\delta=-\frac{\g}{2+\g}.
\eeq
Let us point out that this choice corresponds to the Metropolis dynamics:
\beqa\label{metro}
w_n&=&\min(1,\e^{-\beta\Delta\H})\nonumber\\
&=&\frac{2+\g}{2(1+\g)}
\left(1-\frac{\g}{2+\g}\bigl(\s_n(\s_{n-1}+\s_{n+1})+\s_{n-1}\s_{n+1}\bigr)\right).
\eeqa
Here again, three of the four rates coincide: $w_{+-}=w_{-+}=w_{--}$.
The Metropolis point is located at the intersection of the reversibility line
with the first bisectrix ($\epsr=\delta$).
For $\g=4/5$, as in figure~\ref{reversible}, we thus get $\delta=-2/7$.
At this point, unfortunately,
the quantities whose decay rates can be calculated by the above reduction technique
are the staggered magnetization and the staggered three-spin product~\cite{malte},
which are not relevant to the dynamics of local observables in a ferromagnetic model.

To close, let us notice that the quantum Hamiltonian~(\ref{fham})
does not have any special feature at the KDH and Metropolis points.
In particular it seems to remain non-integrable.
Further considerations on the subtle relationship
between partial solvability and integrability in quantum chains
and related matters can be found in~\cite{prs}.
The explicit examples analyzed there correspond to cases of free fermions.
They are thus analogous to our solvable models.

\section{Discussion}
\label{discussion}

The present work contains a detailed investigation
of the generic dynamics of the ferromagnetic Ising chain introduced
in~\cite{gb09,cg11,cg13}.
This generic one-dimensional kinetic Ising model extends the Glauber model~\cite{glauber}
to a two-parameter space corresponding to non-linearity and irreversibility,
respectively measured by the deformation parameters~$\delta$ and $\eta$.
While the introduction of the non-linearity parameter~$\delta$ is already
present in~\cite{glauber}, the introduction of the irreversibility parameter~$\eta$
is a novelty.
This generic dynamics is also the most general single-spin-flip dynamics which fulfills,
besides global balance with respect to the ferromagnetic
Hamiltonian~(\ref{eq:hamil}),
spin reversal symmetry and a spatially homogeneous dynamical influence
of nearest neighbours only.

The present study has the virtue of organizing many partial results which were
scattered in the literature concerning the role of the non-linearity parameter
$\delta$ for the reversible chain,
and of unveiling novel features when the irreversibility parameter $\eta$ is
simultaneously present.
In view of the extensive body of knowledge on kinetic Ising models, especially
in one dimension, since Glauber's seminal work in 1963 (see e.g.~the reviews
in~\cite{privman}), it may appear surprising that only recently has this
generic kinetic Ising model been considered.

At infinite temperature, where Glauber dynamics accounts for independent spins,
the presence of deformation parameters already has drastic consequences.
The key observable is the overlap $C(0,t)$ between the initial spin
configuration and the current configuration at time $t$.
Along the reversibility line, the non-linearity slows down the dynamics.
The dependence of the relaxation rate of $C(0,t)$ on $\delta$ is given exactly.
An extreme non-linearity parameter ($\delta=-1$) yields a microcanonical dynamics,
where the dynamics of domain walls is described by a SEP.
The stretched exponential relaxation of $C(0,t)$ is also fully characterised.
All along the reversibility line,
an infinitesimal amount $\eta$ of irreversibility
induces an oscillatory relaxation of $C(0,t)$,
with the period of oscillations diverging as $1/\abs{\eta}$.
With irreversibility, the SEP domain wall dynamics becomes an ASEP dynamics,
featuring a stretched exponential relaxation modulated by oscillations.
Investigating the spectrum of the Markov matrices of finite chains
directly yields a clear picture of the main features of the dynamics,
such as reversibility or integrability (see figures~\ref{int} to~\ref{clouds}).

At finite temperature, the main novel feature is
the occurrence of two successive thresholds
in the irreversibility parameter $\eta$.
This phenomenon was first uncovered by analytical means on the solvability line
$\delta=0$~\cite{cg11}.
Beyond a first threshold ($\eta=\pm\eta_c$),
a random and a thermalized initial states yield different spin relaxation times.
This threshold also marks the onset of a strong
violation of the equilibrium fluctuation-dissipation theorem.
Beyond a second threshold ($\eta=\pm\eta_0$),
all two-time correlation functions exhibit an oscillatory relaxation.
Figure~\ref{regimes} summarizes the phase diagram in the $\gamma$--$\eta$ plane for these correlation functions.
An analogous phase diagram is expected to prevail for generic parameter values.
In the present work we restricted the study to the second threshold,
whose presence was demonstrated by simulations.

The qualitative picture sketched above has been complemented
by a good deal of quantitative predictions, coming from a variety of approaches
including numerical simulations, time series expansions,
and the spectra of Markov matrices and quantum spin Hamiltonians.

\ack

It is a pleasure to thank M. Henkel, K. Mallick,
V. Pasquier, S. Prolhac, V. Terras and C. Toninelli for stimulating discussions
on various aspects of this work.

\end{document}